\documentclass[useAMS]{mn2e} \voffset=-1cm \usepackage{times}
\usepackage{psfig} \usepackage{epsfig} \usepackage{amssymb}
\usepackage{amsbsy}

\def\etal{{\it et al.\ }} 
 \def\beqn{\vspace{2mm}
\begin{eqnarray}} \def\eeqn{\vspace{2mm} \end{eqnarray}}
\def\omegal{{\Omega_\Lambda}} \def\omegam{{\Omega_{\rm m}}}
  
\def\sqdeg{{\rm~deg}^2} \def\bj{b_{\rm J}}
\def\gsim{\lower.73ex\hbox{$\sim$}\llap{\raise.4ex\hbox{$>$}}$\,$}
\def\lsim{\lower.73ex\hbox{$\sim$}\llap{\raise.4ex\hbox{$<$}}}

\begin{document}

\title[The 2dFGRS: Wiener Reconstruction of the Cosmic Web]{The 2dF Galaxy 
Redshift Survey: Wiener Reconstruction of the
Cosmic Web}
\author[Erdo\~{g}du \etal] {
\parbox[t]{\textwidth}{Pirin Erdo\~{g}du$^{1,2}$,
Ofer Lahav$^1$,
Saleem Zaroubi$^3$,
George Efstathiou$^1$,
Steve Moody,
John A.\ Peacock$^{12}$,
Matthew Colless$^{17}$,
Ivan K.\ Baldry$^9$,
Carlton M.\ Baugh$^{16}$,
Joss Bland-Hawthorn$^7$,
Terry Bridges$^7$,
Russell Cannon$^7$,
Shaun Cole$^{16}$,
Chris Collins$^4$,
Warrick Couch$^5$,
Gavin Dalton$^{6,15}$,
Roberto De Propris$^{17}$,
Simon P.\ Driver$^{17}$,
Richard S.\ Ellis$^8$,
Carlos S.\ Frenk$^{16}$,
Karl Glazebrook$^9$,
Carole Jackson$^{17}$,
Ian Lewis$^6$,
Stuart Lumsden$^{10}$,
Steve Maddox$^{11}$,
Darren Madgwick$^{13}$,
Peder Norberg$^{14}$,
Bruce A.\ Peterson$^{17}$,
Will Sutherland$^{12}$,
Keith Taylor$^8$
(The 2dFGRS Team)}
\vspace*{6pt} \\
$^1$Institute of Astronomy, Madingley
Road, Cambridge CB3 0HA, UK \\
$^2$Department of Physics, Middle East
Technical University, 06531, Ankara, Turkey \\
$^3$Max Planck Institut
f\"{u}r Astrophysik, Karl-Schwarzschild-Stra{\ss}e 1, 85741 Garching,
Germany \\
$^4$Astrophysics Research Institute, Liverpool John Moores University,
    Twelve Quays House, Birkenhead, L14 1LD, UK \\
$^5$Department of Astrophysics, University of New South Wales, Sydney,
    NSW 2052, Australia \\
$^6$Department of Physics, University of Oxford, Keble Road,
    Oxford OX1 3RH, UK \\
$^7$Anglo-Australian Observatory, P.O.\ Box 296, Epping, NSW 2111,
    Australia\\
$^8$Department of Astronomy, California Institute of Technology,
    Pasadena, CA 91025, USA \\
$^9$Department of Physics \& Astronomy, Johns Hopkins University,
       Baltimore, MD 21118-2686, USA \\
$^{10}$Department of Physics, University of Leeds, Woodhouse Lane,
       Leeds, LS2 9JT, UK \\
$^{11}$School of Physics \& Astronomy, University of Nottingham,
       Nottingham NG7 2RD, UK \\
$^{12}$Institute for Astronomy, University of Edinburgh, Royal Observatory,
       Blackford Hill, Edinburgh EH9 3HJ, UK \\
$^{13}$Lawrence Berkeley National Laboratory, 1 Cyclotron Road,
       Berkeley, CA 94720, USA \\
$^{14}$ETHZ Institut f\"{u}r Astronomie, HPF G3.1, ETH H\"{o}nggerberg, 
CH-8093
       Zurich, Switzerland \\
$^{15}$Rutherford Appleton Laboratory, Chilton, Didcot, OX11 0QX, UK
\\
$^{16}$Department of Physics, University of Durham, South Road, Durham
DH1 3LE, UK \\
$^{17}$Research School of Astronomy \& Astrophysics, The Australian
    National University, Weston Creek, ACT 2611, Australia\\
}

\maketitle

\begin{abstract}
We reconstruct the underlying density field of the 2 degree Field
Galaxy Redshift Survey (2dFGRS) for the redshift range $0.035<z<0.200$ using the
Wiener Filtering method. The Wiener Filter suppresses shot noise
and accounts for selection and incompleteness effects. The method
relies on prior knowledge of the 2dF power spectrum of fluctuations and the
combination of matter density and  bias parameters however the results
are only slightly affected by changes to these parameters. We
present maps of the density field in two different resolutions:
$5h^{-1}$ Mpc and $10h^{-1}$ Mpc. We identify all major
superclusters and voids in the survey. In particular, we find
two large superclusters and two large local voids. A version of this paper with full set of colour maps can be found at
http://www.ast.cam.ac.uk/~pirin.

\end{abstract}


\section{INTRODUCTION}

Historically, redshift surveys have provided the data and the test
ground for much of the research on the nature of clustering and the 
distribution of galaxies. In the past few years, observations of
large scale structure have improved greatly. Today, with the 
development of fibre-fed spectrographs that can  simultaneously
measure spectra of hundreds of galaxies, cosmologists
have at their fingertips large redshift surveys such as 2 degree Field (2dF) and 
Sloan Digital Sky Survey (SDSS). The analysis of these redshift surveys yield
invaluable cosmological information. On the quantitative side, with
the assumption that the galaxy distribution arises from the
gravitational instability of small fluctuations generated in the early
universe, a wide range of statistical measurements can be obtained,
such as the power spectrum and bispectrum. Furthermore, a qualitative
understanding of galaxy distribution provides insight into the mechanisms
of structure formation that generate the complex pattern of sheets and 
filaments comprising the `cosmic web' (Bond, Kofman \& Pogosyan 1996)   we observe and
allows us to map a wide variety of structure, including clusters, 
superclusters and voids.

Today, many more redshifts are available for galaxies than direct distance 
measurements. This discrepancy inspired a great deal
of work on methods for reconstruction of the real-space density field from 
that observed in redshift-space.
These methods use a variety of functional representations (e.g. Cartesian, 
Fourier, spherical harmonics or wavelets)
and smoothing techniques (e.g. a Gaussian sphere or a Wiener Filter).
There are physical as well as practical reasons why one would be
interested in smoothing the observed density field. It is often
assumed that the galaxy distribution samples the underlying smooth
density field and the two are related by a proportionality constant,
the so-called linear bias parameter, $b$. The
finite sampling of the smooth underlying field introduces Poisson
`shot noise' \footnote{Another popular
model for galaxy clustering is the halo model where the linear bias
parameter depends on the mass of the dark matter halos where the
galaxies reside. For this model, the mean number of galaxy pairs in a
given halo is usually lower than the Poisson expectation.}. Any
robust reconstruction technique must reliably mitigate
the statistical uncertainties due to shot noise.
Moreover, in redshift surveys, the actual number of galaxies in a
given volume is larger than the number observed, in particular in
magnitude limited samples where at large distances only very
luminous galaxies can be seen.

In this paper, we analyse large scale structure in the 2 degree Field
Galaxy Redshift Survey (2dFGRS, Colless \etal 2001), which has now
obtained the redshifts for approximately 230,000 galaxies.
We recover the underlying density field, characterised
by an assumed power spectrum of fluctuations, from the observed field which
suffers from incomplete sky coverage (described by the angular mask) and
incomplete galaxy sampling due to its magnitude limit (described by the
selection function).  The filtering is achieved by a Wiener Filter
(Wiener 1949, Press \etal 1992) within the framework of both linear
and non-linear 
theory of density fluctuations.  The Wiener Filter is optimal in the
sense that the variance between the derived reconstruction and the
underlying true density field is minimised. As opposed to {\it ad hoc} 
smoothing schemes,
the smoothing due to the Wiener Filter is determined by the data. In the limit 
of high
signal-to-noise, the Wiener Filter modifies the observed data only
weakly, whereas it suppresses the contribution of the
data contaminated by shot noise.

The Wiener Filtering is a well known technique and has
been applied to many fields in astronomy (see Rybicki \& Press 1992).
For example, the method was used to reconstruct the angular
distribution (Lahav \etal 1994), the real-space density, velocity
and gravitational potential fields of the 1.2 Jy-$IRAS$ (Fisher \etal
1995) and $IRAS$ PSCz surveys (Schmoldt \etal 1999). The Wiener Filter
was also applied to the reconstruction of the 
angular maps of the Cosmic Microwave Background temperature
fluctuations (Bunn \etal 1994, Tegmark \& Efstathiou 1995, Bouchet \&
Gispert 1999). A
detailed formalism of the Wiener 
Filtering method as it pertains to the large scale structure
reconstruction can be found in Zaroubi \etal (1995). 

This paper is structured as follows: we begin with a brief review of
the formalism of the Wiener Filter method.  A summary of 2dFGRS data
set, the survey mask and the selection function are given in Section
3. In section 4, we outline the scheme used to pixelise the survey.
In section 5, we give the formalism for the covariance matrix used in
the analysis. After that, we describe the application
of the Wiener Filter method to 2dFGRS and present detailed maps of the
reconstructed field.  In section 7, we identify the superclusters and
voids in the survey.

Throughout this paper, we assume a $\Lambda$CDM cosmology with $\omegam
= 0.3$ and $\omegal = 0.7$ and $H_0 = 100 h^{-1} {\rm km} {\rm s}^{-1}
{\rm Mpc}^{-1}$.

\section{Wiener Filter}

In this section, we give a brief description of the Wiener Filter
method. For more details, we refer the reader to Zaroubi \etal (1995).
Let's assume that we have a set of measurements, $\{d_\alpha\}\
(\alpha=1,2,\dots N)$ which are a linear convolution of the true
underlying signal, $s_\alpha$, plus a contribution from statistical
noise, $\epsilon_\alpha$, such that
\begin{equation}
d_\alpha = s_\alpha + \epsilon_\alpha.
\label{datavec}
\end{equation}
The Wiener Filter is defined as the {\it linear} combination of the
observed data which is closest to the true signal in a minimum
variance sense. More explicitly, the Wiener Filter estimate,
$s_\alpha^{WF}$, is given by $s_\alpha^{WF} = F_{\alpha\beta}\,
d_\beta$ where the filter is chosen to minimise the variance of the
residual field, $r_\alpha$:
\begin{equation}
\langle |r_\alpha^2|\rangle = \langle
|s^{WF}_{\alpha}-s_\alpha|^2\rangle.  \qquad
\label{vares}
\end{equation}

It is straightforward to show that the Wiener Filter is given by
\begin{equation}
F_{\alpha\beta} = \langle s_\alpha d_\gamma^\dagger \rangle \langle
d_\gamma d_\beta^\dagger \rangle^{-1},\qquad
\label{wienerfilter}
\end{equation}
where the first term on the right hand side is the signal-data
correlation matrix;
\begin{equation}
\langle s_\alpha d_\gamma^\dagger\rangle = \langle s_\alpha
s^\dagger_\gamma\rangle,\qquad
\end{equation}
and the second term is the data-data correlation matrix;
\begin{equation}
\langle d_\alpha d_\beta^\dagger\rangle = \langle s_\gamma
s^\dagger_\delta\rangle + \langle \epsilon_\alpha
\epsilon_\beta^\dagger \rangle. \qquad
\label{ddcor}
\end{equation}

In the above equations, we have assumed that the signal and noise are
uncorrelated. From equations~\ref{wienerfilter} and ~\ref{ddcor}, it is clear that, in order to
implement the Wiener Filter, one must construct a {\it prior} which depends 
on the mean of the signal (which is 0 by construction) and the variance
of the signal and noise. The assumption of a prior may be alarming at
first glance. However, slightly inaccurate values of Wiener Filter will only introduce second order errors to the full reconstruction (see
Rybicki \& Press 1992).  
The dependence of the Wiener Filter on the prior can be made clear by 
defining
signal and noise matrices as $C_{\alpha\beta}=\langle s_\alpha
s^\dagger_\beta\rangle$ and $N_{\alpha\beta}=\langle \epsilon_\alpha
\epsilon^\dagger_\beta\rangle$. With this notation, we can rewrite the
equations above so that ${\bf s}^{WF}$ is given as
\begin{equation}
{\bf s}^{WF} = {\bf C} \left[ {\bf C} + {\bf N}\right]^{-1} {\bf
d}. \qquad
\label{swfvec}
\end{equation}
The mean square residual given in equation~\ref{vares} can then be
calculated as
\begin{equation}
\langle {\bf r} {\bf r}^\dagger\rangle = {\bf C} \left[ {\bf C} + {\bf
N}\right]^{-1} {\bf N}. \qquad
\label{res}
\end{equation}

Formulated in this way, we see that the purpose of the Wiener Filter is to
attenuate the contribution of low signal-to-noise ratio data. The
derivation of the Wiener Filter given above follows from the sole 
requirement of
minimum variance and requires only a model for the variance of the
signal and noise. The Wiener Filter can also be derived using the laws of
conditional probability if the underlying distribution functions for
the signal and noise are assumed to be Gaussian. For the Gaussian
prior, the Wiener Filter estimate is both the maximum {\it posterior} 
estimate and the mean field (see Zaroubi \etal 1995).

As several authors point out (e.g. Rybicki \& Press 1992, Zaroubi
2002), the Wiener Filter is a biased estimator since it predicts a null
field in the absence of good data, unless the field itself has zero
mean. Since we constructed the density field to have zero mean, we are
not worried about this bias. However, the observed field 
deviates from zero due to selection effects and so one needs to be
aware of this bias in the reconstructions.

It is well known that the peculiar velocities of galaxies distort
clustering pattern in redshift space. On small scales, the random
peculiar velocity of each galaxy causes smearing along the line of
sight, known as {\it the finger of God}. On larger scales, there is
compression of structures along the line of sight due to coherent
infall velocities of large-scale structure induced by gravity.  One of
the major difficulties in analysing redshift surveys is the
transformation of the position of galaxies from redshift space to real
space. For all sky surveys, this issue can be addressed using several
methods, for example the iterative method of Yahil \etal (1991) and
modified Poisson equation of Nusser \& Davis 1994. However, these
methods are not applicable to surveys which are not all sky as they
assume that in linear theory, the peculiar velocity of any galaxy is a
result of the matter distribution around it, and the gravitational field
is dominated by the matter distribution inside the volume of the
survey. For a survey like 2dFGRS, within the limitation of linear
theory where the redshift space density is a linear transformation of
the real space density, a Wiener Filter can be used to transform from 
redshift
space to real space (see Fisher \etal (1995) and Zaroubi \etal (1995)
for further details).  This can be written as
\begin{equation}
s^{WF}(r_\alpha) = \langle s(r_\alpha) d(s_\gamma) \rangle \langle
d(s_\gamma) d(s_\beta) \rangle^{-1} d(s_\beta), \qquad
\label{opt}
\end{equation}
where the first term on the right hand side is the cross-correlation
matrix of real and redshift space densities and {{\bf s}({\bf r})} is the
position vector in redshift space.  It is worth emphasising that this
method is limited as it only recovers the peculiar velocity field
generated by the mass sources represented by the galaxies within the
survey boundaries. It does not account for possible external forces.
This limitation can only be overcome by comparing the 2dF survey with
all sky surveys.

\section{The Data}

\subsection{The 2dFGRS data}

The 2dFGRS, now completed, is selected in the photometric $\bj$\ band
from the APM
galaxy survey (Maddox, Efstathiou \& Sutherland 1990) and its
subsequent extensions (Maddox \etal 2003, in preparation). The survey
covers about $2000\sqdeg$ and is made up of two declination strips,
one in the South Galactic Pole region (SGP) covering approximately
$-37^\circ\negthinspace.5<\delta<-22^\circ\negthinspace.5$,
$-35^\circ\negthinspace.0<\alpha<55^\circ\negthinspace.0$ and the
other in the direction of the North Galactic Pole (NGP), spanning
$-7^\circ\negthinspace.5<\delta<2^\circ\negthinspace.5$,
$147^\circ\negthinspace.5<\alpha<222^\circ\negthinspace.5$.
In addition to these contiguous regions, there are a number of
randomly located circular 2-degree fields scattered over the full
extent of the low extinction regions of the southern APM galaxy
survey.

The magnitude limit at the start of the survey was set at $\bj =
19.45$ but both the photometry of the input catalogue and the dust
extinction map have been revised since and so there are small
variations in magnitude limit as a function of position over the sky
which are taken into account using the magnitude limit mask. The
effective median magnitude limit, over the area of the survey, is $\bj
\approx 19.3$ (Colless \etal 2001). 

We use the data obtained prior to May 2002, when the survey was nearly
complete. This includes 221\,283 unique, reliable galaxy
redshifts. We analyse a magnitude-limited sample with redshift limits
$z_{\rm{min}} = 0.035$ and $z_{\rm{max}} = 0.20$.  The median redshift
is $z_{\rm med} \approx 0.11$.  We use 167\,305 galaxies in total,
98\,129 in the SGP and 69\,176 in the NGP. We do not include the
random fields in our analysis. 

The 2dFGRS database and full documentation are available on the WWW at
http://www.mso.anu.edu.au/2dFGRS/.

\subsection{The Mask and The Radial Selection Function of 2dFGRS}
The completeness of the survey varies according to the position in the
sky due to unobserved fields, particularly at the survey edges, and
unfibred objects in the observed fields because of collision
constraints or broken fibres.

\begin{figure*}
\psfig{figure=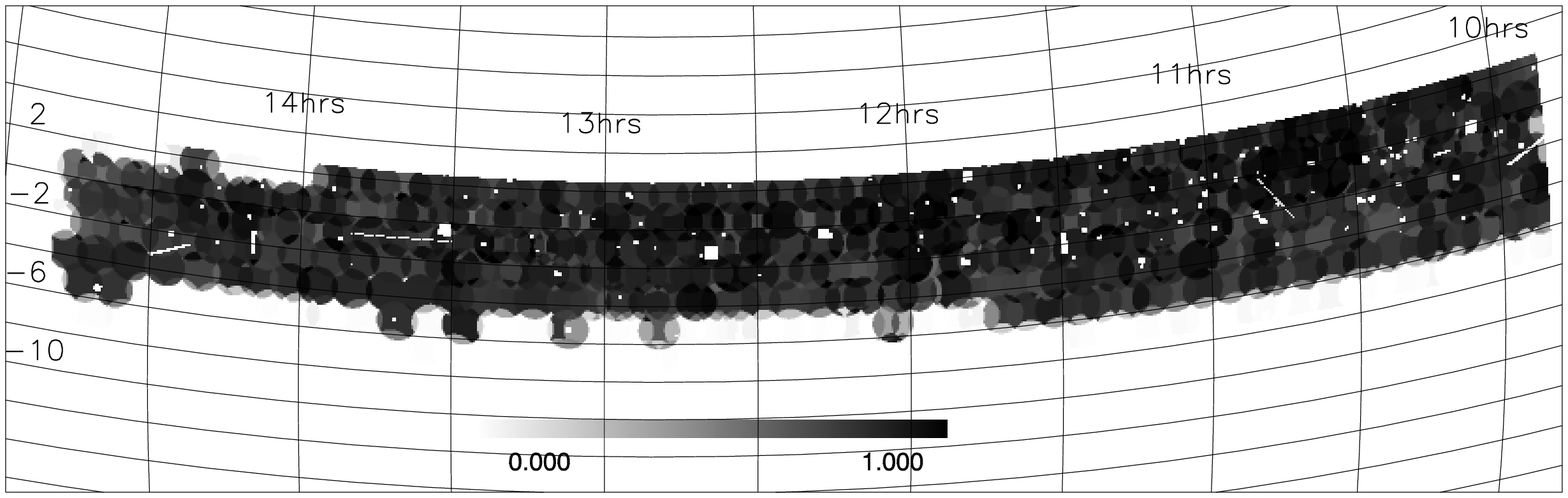,angle=0,width=\textwidth,clip=}
\psfig{figure=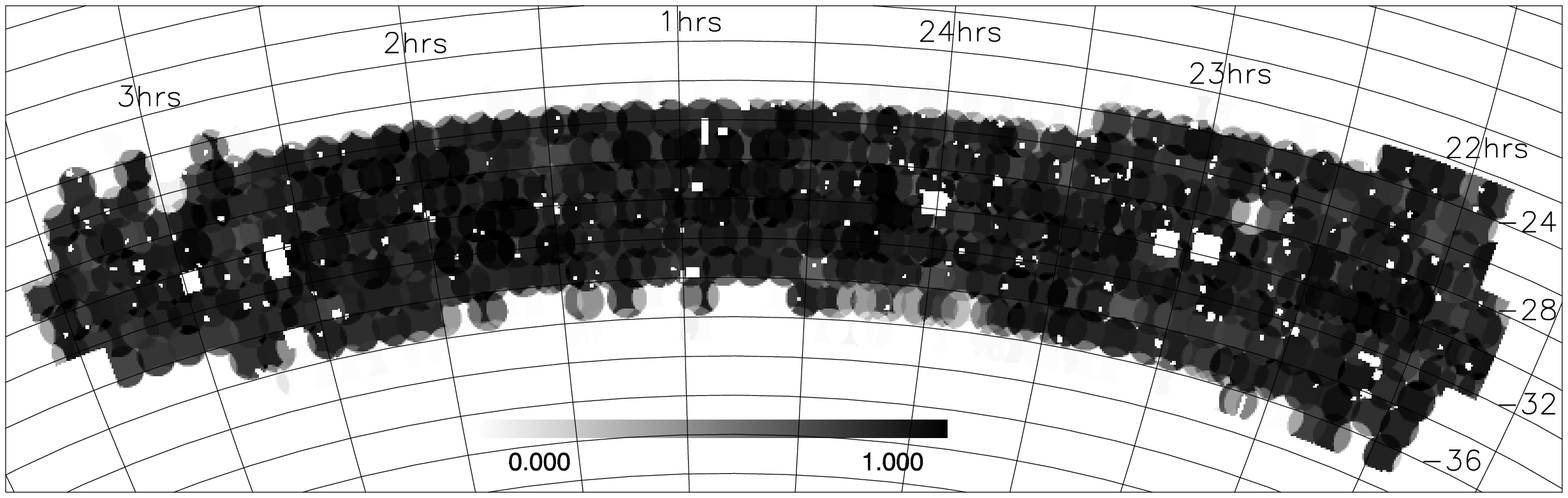,angle=0,width=\textwidth,clip=}
\caption[] {The redshift completeness masks for the NGP (top) and
SGP (bottom) in Equatorial coordinates. The gray scale shows the completeness fraction.}
\label{mask}
\end{figure*}

For our analysis, we make use of two different masks (Colless
\etal~2001; Norberg \etal~2002). The first of these masks is the
redshift completeness mask defined as the ratio of the number of
galaxies for which redshifts have been measured to the total number of
objects in the parent catalogue. This spatial incompleteness is
illustrated in Figure~\ref{mask}. The second mask is the magnitude
limit mask which gives the extinction corrected magnitude limit of the
survey at each position.

The radial selection function
gives the probability of observing a galaxy for a given redshift and
can be readily calculated from the galaxy luminosity function:

\begin{equation}
\Phi(L)dL =
\Phi^*\left(\frac{L}{L^*}\right)^\alpha\exp\left(-\frac{L}{L^*}\right)\frac{dL}{L^*}
\;.
\end{equation}
where for the concordance model: $\alpha =
-1.21\pm0.03$, $\log_{10}L_* = -0.4(-19.66\pm0.07+ 5 \log_{10}(h))$
and $\Phi_* = 0.0161\pm0.0008 h^3 $ (Norberg \etal 2002).

The selection function can then be expressed as
\begin{equation}
\phi(r) = \frac {\int_{L(r)}^{\infty}\Phi(L)dL}
{\int_{L_{min}}^{\infty}\Phi(L)dL},
\end{equation}
where $L(r)$ is the minimum luminosity detectable at luminosity
distance $r$ (assuming the concordance model), evaluated for the concordance 
model, $L_{\rm min}={\rm
Min}(L(r),L_{\rm com})$ and $L_{\rm com}$ is the minimum luminosity
for which the catalogue is complete and varies as a function of
position over the sky. For distances considered in this paper, where
the deviations from the Hubble flow are relatively small, the
selection function can be approximated as
$\phi(r)\approx\phi(z_{gal})$.
Each galaxy, $gal$, is then assigned the weight
\begin{equation}
w(gal) ={ 1 \over \phi(z_{gal}) M(\Omega_{i})}
\label{weight}
\end{equation}
where $\phi(z_{gal})$ and $M(\Omega_i)$ are the values of the
selection function for each galaxy and angular survey mask for each
cell $i$ (see Section 4), respectively.

\section{Survey Pixelisation}

In order to form a data vector of overdensities, the survey needs to
be pixelised.  There are many ways to pixelise a survey: equal sized
cubes in redshift space, igloo cells, spherical harmonics, Delauney
tessellation methods, wavelet decomposition, etc. Each of these
methods have their own advantages and disadvantages and they should be
treated with care as they form functional bases in which all the
statistical and physical properties of cosmic fields are retained.

\begin{figure}
\psfig{figure=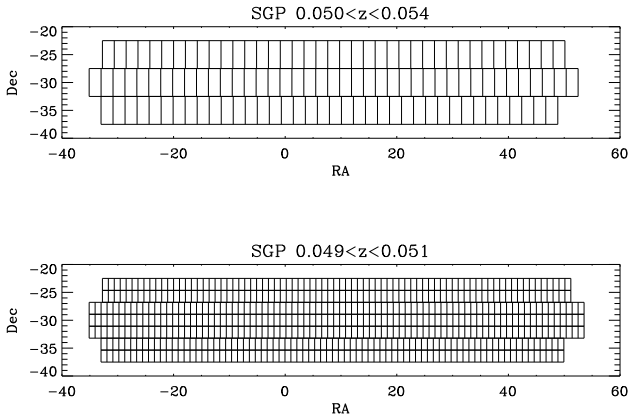,angle=0,width=0.5\textwidth,clip=}
\caption[] {An illustration of the survey pixelisation scheme used in
the analysis, for $10h^{-1}$ Mpc (top) and $5h^{-1}$ Mpc (bottom)
target cells widths. The redshift ranges given on top of each plot.}
\label{pixels}
\end{figure}

The pixelisation scheme used in this analysis is an `igloo' grid with
wedge shaped pixels bounded in right ascension, declination and in
redshift. The pixelisation is constructed to keep the average number
density per pixel approximately constant.  The advantage of using this
pixelisation is that the number of pixels is minimised since the pixel
volume is increased with redshift to counteract the decrease in the
selection function. This is achieved by selecting a `target cell width'
for cells at the mean redshift of the survey and deriving the rest of
the bin widths so as to match the shape of the selection function. The
target cell widths used in this analysis are $10h^{-1}$ Mpc and $5h^{-1}$ Mpc. Once the redshift binning has been
calculated each radial bin is split into declination bands and then
each band in declination is further divided into cells in right
ascension. The process is designed so as to make the cells roughly
cubical. In Figure~\ref{pixels}, we show an illustration of this by
plotting the cells in right ascension and declination for a given
redshift strip.

Although advantageous in many ways, the pixelisation scheme used in
this paper may complicate the interpretation of the reconstructed
field. By definition, the Wiener Filter signal will approach to zero at 
the edges of the survey where the shot noise may dominate. This means the true signal
will be constructed in a non-uniform manner. This effect will be
amplified as the cell sizes get bigger at higher redshifts. Hence,
both of these effects must be considered when interpreting the
results.

\section{Estimating The Signal-Signal Correlation Matrix Over Pixels}

The signal covariance matrix can be accurately modelled by an
analytical approximation (Moody 2003). The calculation of the
covariance matrix is similar to the analysis described by Efstathiou
and Moody (2001) apart from the modification due to
three-dimensionality of the survey.  The covariance matrix for the
`noise free' density fluctuations is $\langle C_{ij} \rangle =\langle
\delta_{i}\delta_{j} \rangle$, where $\delta_{i}$ =
$(\rho_{i}-\bar{\rho})/\bar{\rho}$ in the $i$th pixel. It is estimated by
first considering a pair of pixels with
volumes $V_i$ and $V_j$, separated by distance $\bf r$ so that,

\beqn \langle C_{ij} \rangle & = & \left\langle \frac{1}{V_{i}V_{j}}
\int_{Cell_{i}}^{} \delta({\bf x}) dV_{i} \int_{Cell_{j}}^{}
\delta({\bf x+r}) dV_{j} \right\rangle \\ & = &
\frac{1}{V_{i}V_{j}}\int_{Cell_{i}}^{}\int_{Cell_{j}}^{} \langle
\delta({\bf x})\delta({\bf x+r}) \rangle dV_{i}dV_{j} \\ & = &
\frac{1}{V_{i}V_{j}} \int_{Cell_{i}}^{}\int_{Cell_{j}}^{}\xi({\bf r})
dV_{i}dV_{j} \eeqn where the isotropic two point correlation function
$\xi(r)$ is given by \beqn \xi(r) &=&
\frac{1}{(2\pi)^{3}}\int_{}^{}P(k)e^{-i{\bf k\cdot r}}d^{3}k, \eeqn
and therefore, \beqn \langle C_{ij}\rangle &=&
\frac{1}{(2\pi)^{3}V_{i}V_{j}} \int_{} P(k)d^{3}k \nonumber \\
&\times& \int_{Cell_{i}}^{}\int_{Cell_{j}}^{} e^{-i {\bf k}({\bf
r}_i-{\bf r}_j)} dV_{i} dV_{j}.  \eeqn After performing the Fourier
transform, this equation can be written as \beqn \langle C_{ij}\rangle
& = & \frac{1}{(2\pi)^{3}}\int_{}P(k)S({\bf k},{\bf L}_i)S({\bf
k},{\bf L}_j) C({\bf k},{\bf r})d^{3}k,
\label{integrate}
\eeqn where the functions $S$ and $C$ are given by, \beqn S({\bf
k},{\bf L}) & = & {\rm sinc}(k_{x}L_{x}/2){\rm sinc}(k_{y}L_{y}/2)
{\rm sinc}(k_{z}L_{z}/2)\\ C({\bf k},{\bf r}) & = & \cos(k_{x}r_{x})
\cos(k_{y}r_{y}) \cos(k_{z}r_{z}), \eeqn where the label {\bf L}
describes the dimensions of the cell ($L_{x}$, $L_{y}$, $L_{z}$), the
components of {\bf r} describes the separation between cell centres, {\bf
k}= ($k_{x}$, $k_{y}$, $k_{z}$) is the wavevector and $ {\rm sinc}(x) =
\frac{\sin(x)}{x}$.
The wavevector, $\bf k$ is written in spherical co-ordinates
$k,\theta, \phi$ to simplify the evaluation of $C$.  We define,
\beqn k_{x} &=& k \; \sin(\phi)\cos(\theta)\\ k_{y} &=& k \;
\sin(\phi)\sin(\theta)\\ k_{z} &=& k \; \cos(\phi).
\eeqn. Equation~\ref{integrate} can now be integrated over $\theta$ and $\phi$
to form the kernel $G_{ij}(k)$ where, \beqn
G_{ij}(k)=\frac{1}{\pi^{3}}\int_{0}^{\pi/2}\int_{0}^{\pi/2} S({\bf
k},{\bf L}_i)S({\bf k},{\bf L}_j) C({\bf k},{\bf r}) \sin(\phi)d\theta
d\phi, \eeqn so that, \beqn \langle C_{ij} \rangle=\int_{}
P(k)G_{ij}(k) k^{2} dk.  \eeqn In practice we evaluate, \beqn \langle
C_{ij} \rangle=\sum_{k} P_{k}G_{ijk}, \eeqn where $P_{k}$ is the
binned bandpower spectrum and $G_{ijk}$ is, \beqn
G_{ijk}=\int_{k_{min}}^{k_{max}}G_{ij}(k){k^2}dk, \eeqn where the
integral extends over the band corresponding to the band power
$P_{k}$.

For cells that are separated by a distance much larger than the cell
dimensions the cell window functions can be ignored, simplifying the
calculation so that, \beqn
G_{ijk}=\frac{1}{(2\pi)^{3}}\int_{k_{min}}^{k_{max}}\mbox{sinc}(kr)
4\pi k^{2}dk, \eeqn where $r$ is the separation between cell centres.

\section{The Application}

\subsection{Reconstruction Using Linear Theory}

In order to calculate the data vector ${\bf d}$ in equation~\ref{swfvec},
we first estimate the number of galaxies 
$N_{i}$ in each pixel $i$: 
\beqn N_{i}=\sum\limits_{gal}^{N_{gal}(i)}\,
w(gal), \eeqn 
where the sum is over all the observed galaxies in the pixel and $w(gal)$ is the weight assigned to each galaxy
(equation~\ref{weight}). The boundaries of each pixel 
are defined by the scheme described in Section 4, using a
target cell width of $10 h^{-1}$ Mpc.
The mean number of galaxies in pixel $i$ is \beqn \bar{N}_{i}= \bar{n}
V_{i}, \eeqn where $V_{i}$ is the volume of the pixel and the mean
galaxy density, $\bar{n}$, is estimated using the equation:
\beqn 
\bar{n} = \frac{\sum\limits_{gal}^{N_{total}}\, w(gal)}
{\int_{0}^{\infty}\mbox dr r^2 \phi(r) w(r)}, 
\eeqn 
where the sum is now over
all the galaxies in the survey. We note that the value for $\bar{n}$
obtained using 
the equation above is consistent with the maximum estimator method
proposed by Davis and Huchra (1982). Using these definitions, we write
the $i$th component of the
data vector ${\bf d}$ as: 
\beqn 
d_{i} = \frac{N_{i}-\bar{N}_{i}}{\bar{N}_{i}}.  
\eeqn
Note that, the mean value of ${\bf d}$ is zero by construction. 

Reconstruction of the underlying signal given in
equation~\ref{swfvec} also requires the signal-signal and the inverse
of the data-data correlation matrices.
The data-data correlation matrix (equation~\ref{ddcor}) is the
sum of noise-noise 
correlation matrix $\cal {\bf N}$ and the signal-signal correlation
matrix ${\bf C}$ formulated in the previous section.  The only change
made is to the calculation of ${\bf C}$ where the real space
correlation function $\xi(r)$ is now 
multiplied by Kaiser's factor in order to correct for the redshift
distortions on large scales. So 
\begin{equation}
\xi_s = {1 \over (2\pi)^3}\int P^{S}(k) \exp(i {\bf k\cdot (r_2-r_1)})
d^3k,
\end{equation}
where $P^{S}(k)$ is the galaxy power spectrum in redshift space, 
\beqn
P^{S}(k)= K[\beta] P^{R}(k),
\label{linpz}
\eeqn 
derived in linear theory. The subscripts $R$ and $S$ in this equation
(and hereafter) denote 
real and redshift space, respectively.
\begin{equation}
K[\beta] = 1 + {\frac{2}{3}} \beta + {\frac {1}{5}} \beta^2
\end{equation}
is the direction averaged Kaiser's (1987) factor, derived using
distant observer 
approximation and with the assumption that the data subtends a small
solid angle with respect to the observer (the latter assumption is
valid for the 2dFGRS but does not hold for a wide angle survey, see
Zaroubi and 
Hoffman, 1996 for a full discussion). 
Equation~\ref{linpz} shows that in order to apply the Wiener Filter
method, we need a model 
for the galaxy power spectrum in redshift-space 
which depends on the real-space power spectrum spectrum and on the
redshift distortion 
parameter, $\beta\equiv\omegam^{0.6}/b$. 

The real-space galaxy power
spectrum is well described by a scale invariant Cold Dark Matter power 
spectrum with shape parameter, $\Gamma$ for the scales concerned in this
analysis. For $\Gamma$, we use the 
value derived from the 2dF survey by   
 Percival \etal (2001) who fitted the
2dFGRS power spectrum over the range of linear scales using the fitting
formulae of Eisenstein and Hu (1998).  Assuming a Gaussian prior on
the Hubble constant $h=0.7\pm0.07$ (based on Freedman \etal 2001),
they find $\Gamma = 0.2\pm0.03$. The normalisation of the power spectrum
is conventionally expressed in terms of the variance of the density
field in spheres of $8 h^{-1} $ Mpc, $\sigma_{8}$. Lahav \etal (2002) use
2dFGRS data to deduce $\sigma_{8{\rm g}}^{S}(L_s,
z_s)=0.94\pm0.02$ for the galaxies in redshift space, assuming
$h=0.7\pm0.07$ at $z_s\approx0.17$ and $L_s\approx1.9L^*$.  We convert
this result to real space using the following equation:
\begin{equation}
\sigma_{8{\rm g}}^{R}(L_s, z_s) = \sigma_{8{\rm
g}}^{S}(L_s,z_s)/K^{1/2}[\beta(L_s, z_s)] \; \;.
\end{equation}
where $K[\beta]$ is Kaiser's factor. 
For our analysis, we need to use
$\sigma_8$ evaluated at the mean redshift of the survey for galaxies
with luminosity $L^*$. However, one needs to assume a model for the
evolution of galaxy clustering in order to find $\sigma_8$ at
different redshifts. Moreover, the conversion from $L_s$ to $L^*$
introduces uncertainties in the calculation. Therefore, we choose 
an approximate value, $\sigma_{8{\rm g}}^{R}\approx0.8$ to normalise
the power spectrum. 
For $\beta$, we adopt the value
found by Hawkins \etal (2002), $\beta(L_s, z_s)=0.49\pm0.09$ which is
estimated at the effective luminosity, $L_s\approx1.4L^*$, and the
effective redshift, $z_s\approx0.15$, of the survey sample. Our
results are not sensitive to minor changes in $\sigma_8$ and $\beta$.

The other component of the data-data correlation matrix is the noise
correlation matrix $\cal {\bf N}$.
Assuming that the noise in different cells is not correlated, the
only non-zero terms in $\cal {\bf N}$ are the
diagonal terms given by the variance - the second central moment - of
the density error in each cell: \beqn {\cal N}_{ii} = \frac
{1}{\bar{N}_{i}^2} \sum\limits_{gal}^{N_{gal}(i)}\, w^2(gal).  \eeqn

The final aspect of the analysis is the reconstruction of the
real-space density field from the redshift-space observations.
This is achieved using
equation~\ref{opt}.  Following Kaiser (1987), using distant observer
and small-angle approximation, the cross-correlation matrix in
equation~\ref{opt} for the linear regime can be written as
\begin{equation}
\langle s({\bf r}) d({\bf s})\rangle = \langle \delta_{\bf r}
\delta_{\bf s} \rangle = \xi(r)(1+\frac{1}{3}\beta),
\label{linsinr}
\end{equation}
where {\bf s} and {\bf r} are position vectors in redshift and real
space, respectively. The term, $(1+\frac{1}{3}\beta)$, is easily
obtained by integrating the direction dependent density field in
redshift space. Using equation~\ref{linsinr}, the transformation from
redshift space to real space simplifies to:
\begin{equation}
s^{WF}({\bf r}) = \frac{1 + {\frac{1}{3}}\beta}{K[\beta]}{\bf
C}\left[K[\beta]{\bf C} + {\cal \bf N}\right]^{-1} {\bf d}. \qquad
\label{appopt}
\end{equation}
As mentioned earlier, the equation above is calculated for linear
scales only and hence small scale distortions (i.e. {\it fingers of
God}) are not corrected for. For this reason, we collapse in redshift
space the fingers seen in 2dF groups (Eke \etal 2003) with more than 75 members, 25
groups in total (11 in NGP and 14 in SGP). All the galaxies in these
groups are assigned the same coordinates. As expected, correcting
these small scale distortions does not change the constructed fields
substantially as these distortions are practically smoothed out
because of the cell size used in binning the data.

\begin{figure*}
\psfig{figure=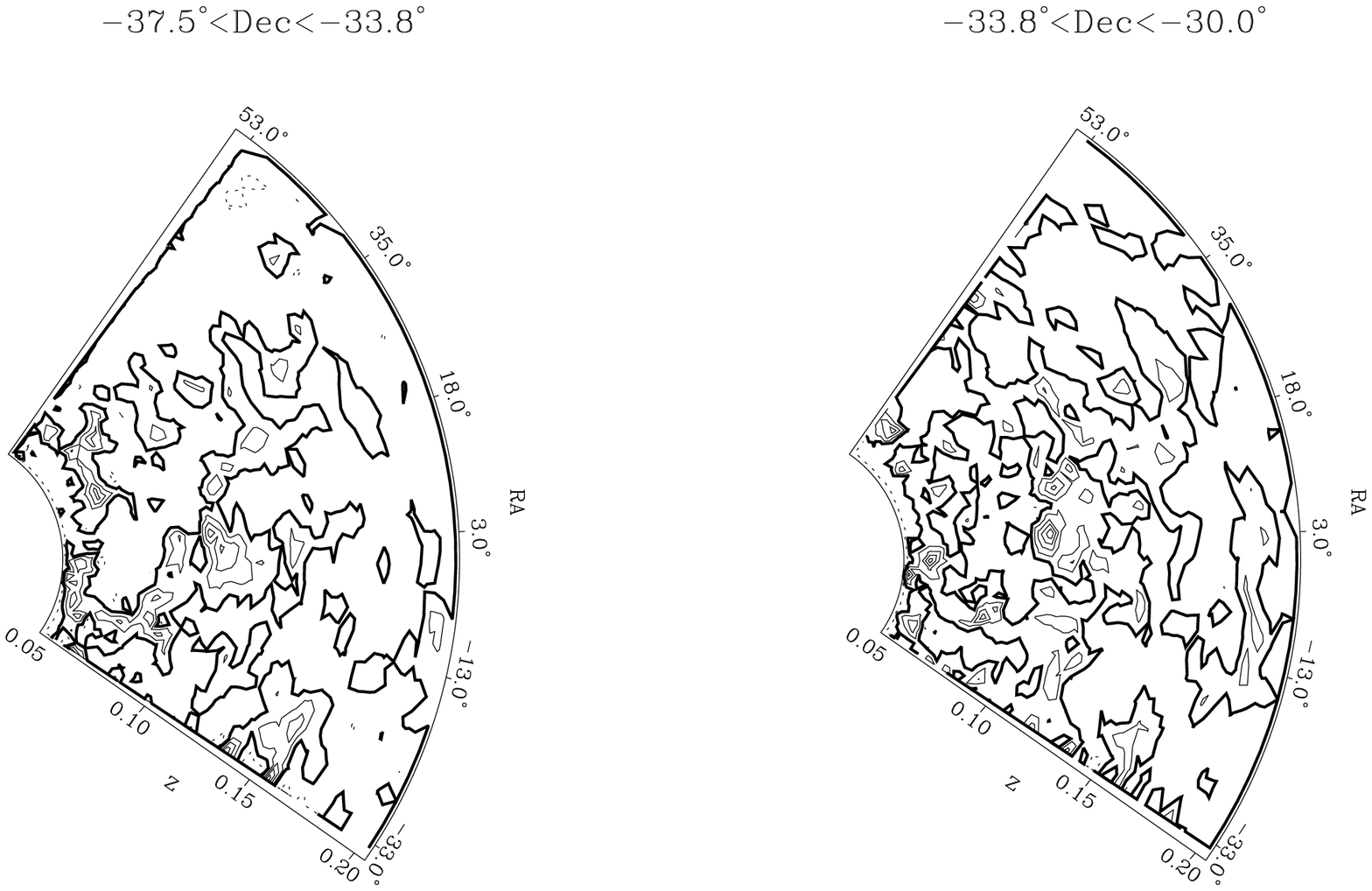,angle=90, width=0.8\textwidth, clip=} 
\psfig{figure=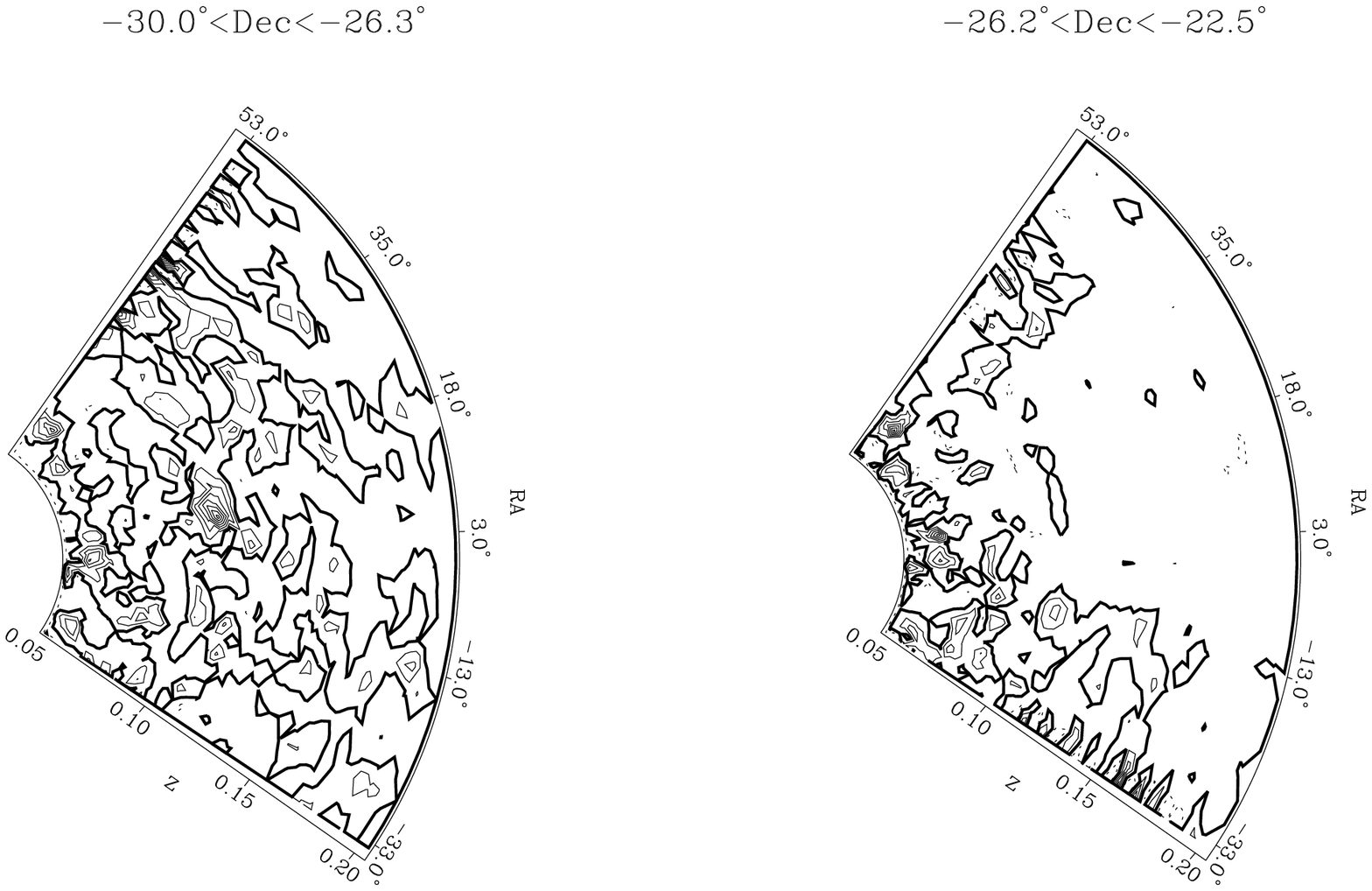,angle=90, width=0.8\textwidth, clip=}
\caption[] {Reconstructions  of the 2dFGRS SGP region in slices of
declination for $10 h^{-1}$
Mpc target cell size . The
declination 
range is given on top each plot. The contours are spaced
at $\Delta\delta = 1.0$ with solid (dashed) lines denoting positive
(negative) contours; the heavy solid contours correspond to $\delta =
0$}
\label{1kelebek}
\end{figure*}

\begin{figure*}
\psfig{figure=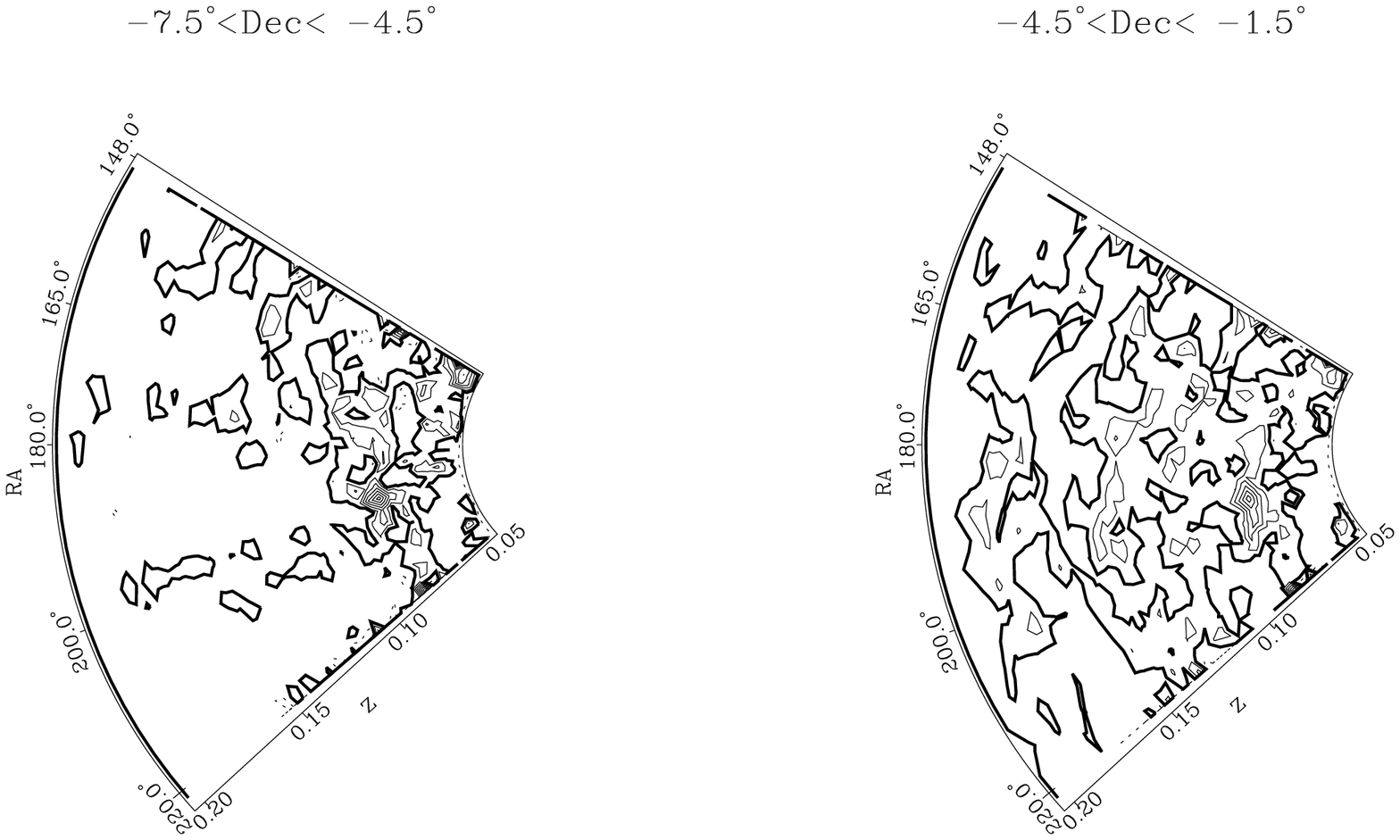,angle=90,width=0.8\textwidth, clip=}
\psfig{figure=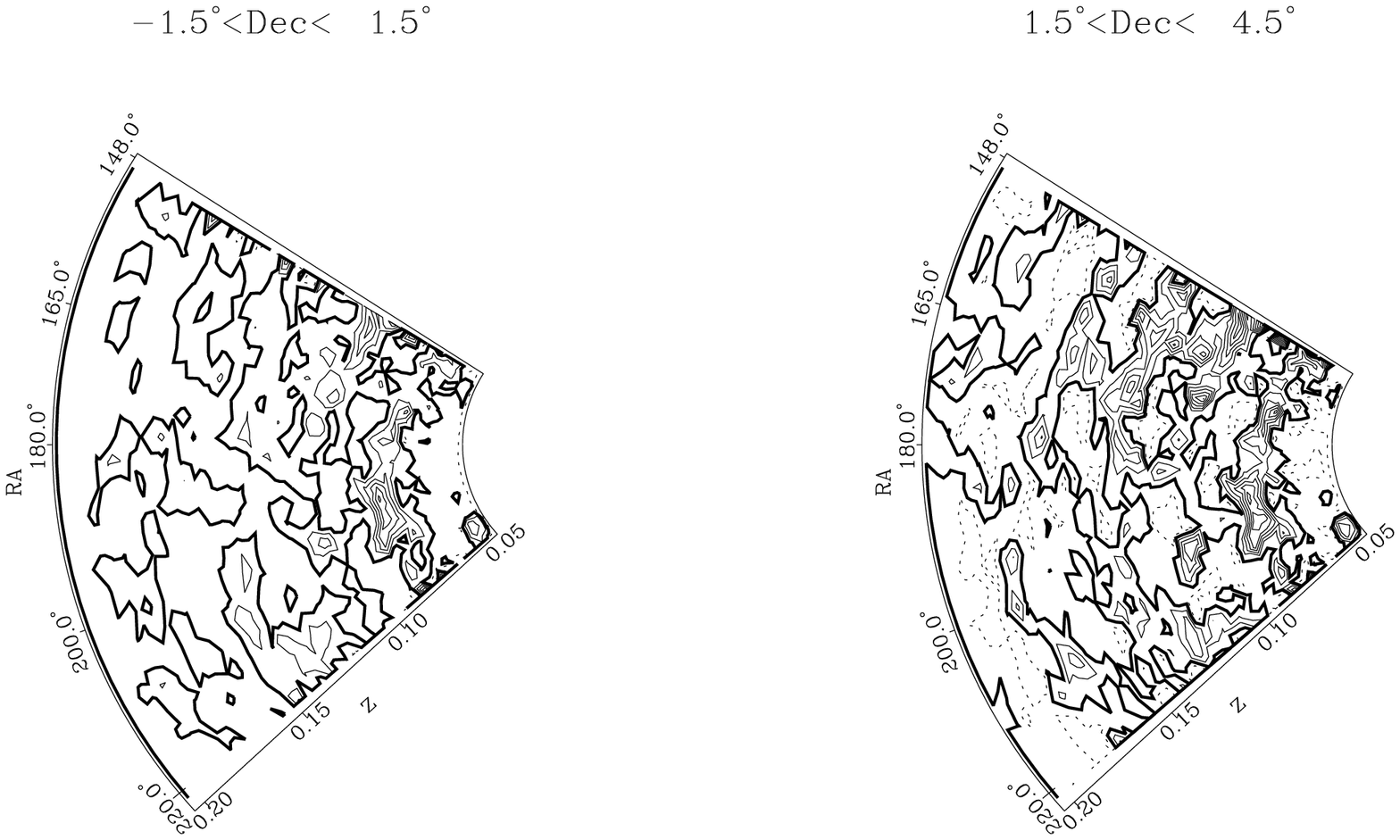,angle=90,width=0.8\textwidth, clip=} 
\caption[] {Reconstructions of the 2dFGRS NGP region in slices of
declination for $10 h^{-1}$
Mpc target cell size.  The
declination 
range is given on top each plot. The contours are spaced
at $\Delta\delta = 1.0$ with solid (dashed) lines denoting positive
(negative) contours; the heavy solid contours correspond to $\delta =
0$}
\label{2kelebek}
\end{figure*}

The maps shown in this section were derived by the technique detailed
above. There are 80 sets of plots which show the density fields as 
strips in $RA$ and $Dec$, 40 maps for SGP and 40 maps for
NGP. Here we just show some examples, the rest of the plots can be
found in url: http://www.ast.cam.ac.uk/$\sim$pirin. For comparison, the top
plots of Figures~\ref{02sgp}, ~\ref{06sgp},
~\ref{09ngp}, and ~\ref{14ngp} show the redshift space density field weighted by
the selection function and the angular mask. The contours are spaced
at $\Delta\delta = 0.5$ with solid (dashed) lines denoting positive
(negative) contours; the heavy solid contours correspond to $\delta =
0$.  Also plotted for comparison are the galaxies (red dots) and the
groups  with $N_{gr}$ number of members (Eke \etal 2003) and 
$9 \leq N_{gr} \leq 17$ (green circles), $18 \leq N_{gr} \leq 44$
(blue squares) and $45 \leq N_{gr}$ (magenta stars). We also show the
number of Abell, APM and EDCC clusters studied by De Propris \etal
(2002) (black upside down triangles). The middle plots show the
redshift space density shown in top plots after the Wiener Filter applied. 
As
expected, the Wiener Filter suppresses the noise. The smoothing performed by the 
Wiener Filter
is variable and increases with distance.  The bottom plots show the
reconstructed real density field ${\bf s}^{WF}({\bf r})$, after correcting
for the redshift distortions. Here the amplitude of density contrast
is reduced slightly. We also plot the
reconstructed fields in declination slices. These plots are shown
in Figures~\ref{1kelebek} and~\ref{2kelebek}.

\begin{figure*}
$\begin{array}{cc}
\psfig{figure=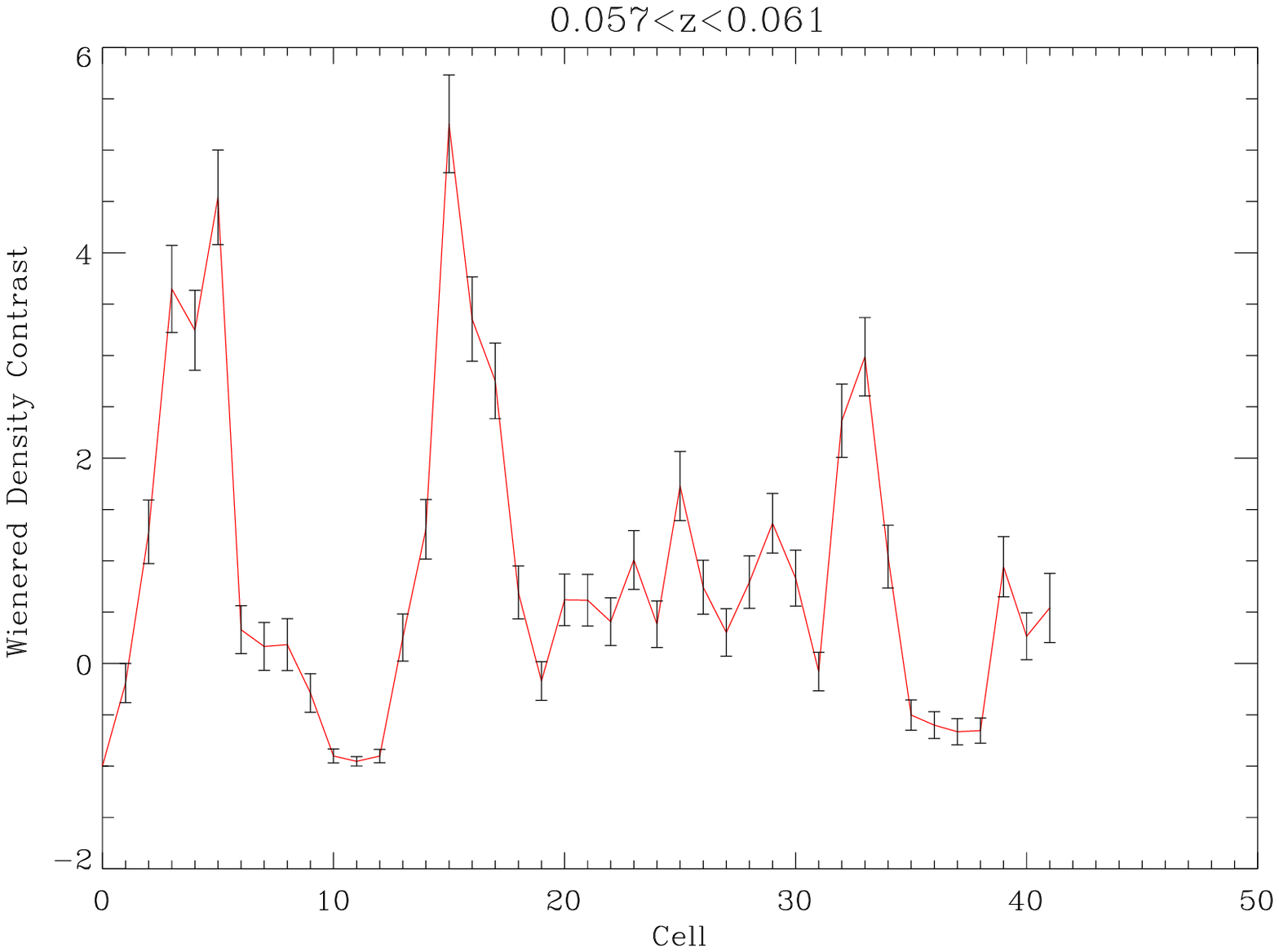,angle=0,width=0.25\textwidth, clip=} &
\psfig{figure=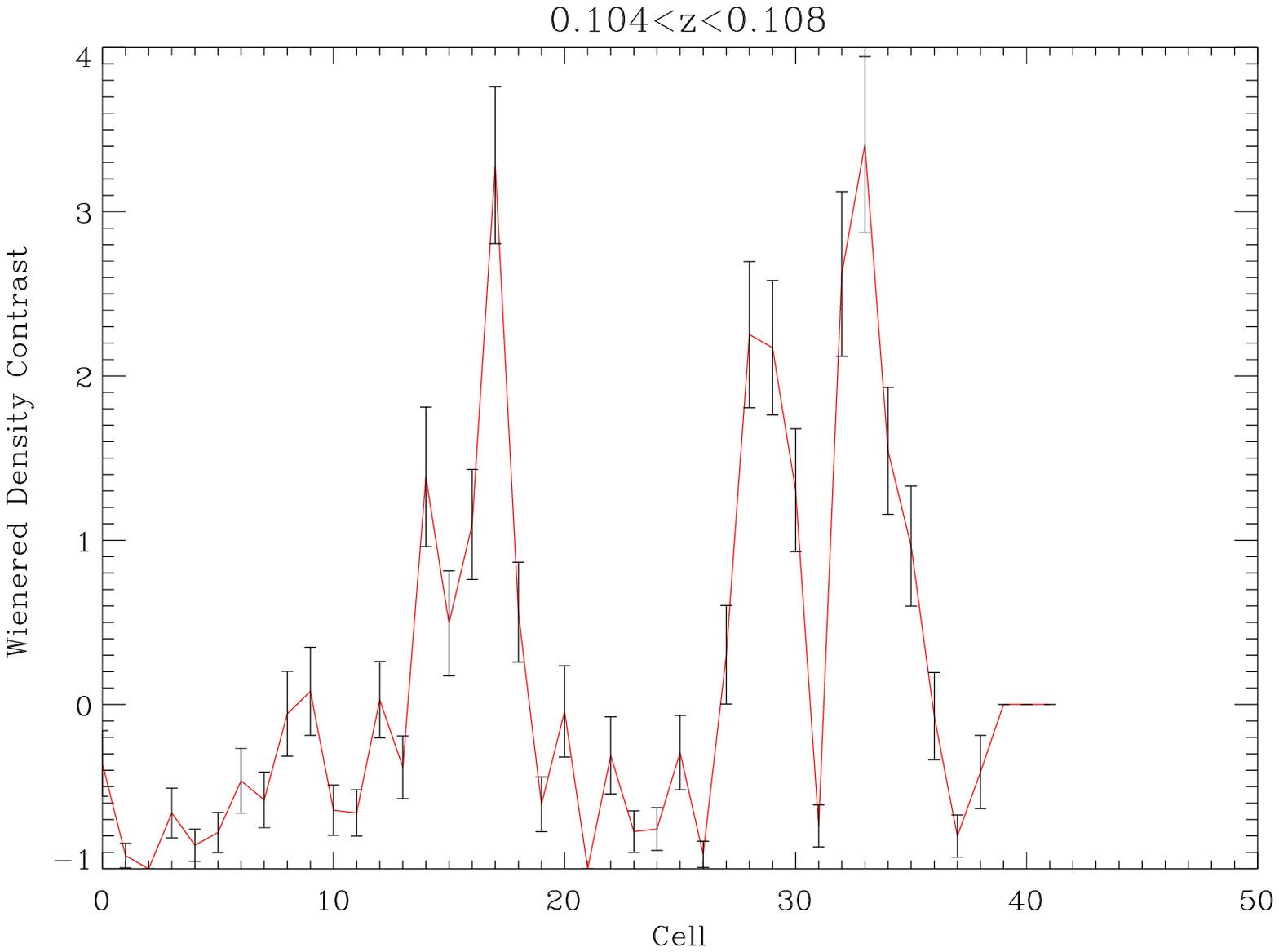,angle=0,width=0.25\textwidth, clip=} \\ 
\end{array}$
\caption[] {The plot of overdensities in SGP for the each redshift
slice for  $10 h^{-1}$
Mpc target cell size shown above. Also plotted are the variances of
the residual 
associated for each cell.The increase in cell number indicates the
increase $RA$ in each redshift slice.}
\label{1res}
\end{figure*}

\begin{figure*}
$\begin{array}{cc}
\psfig{figure=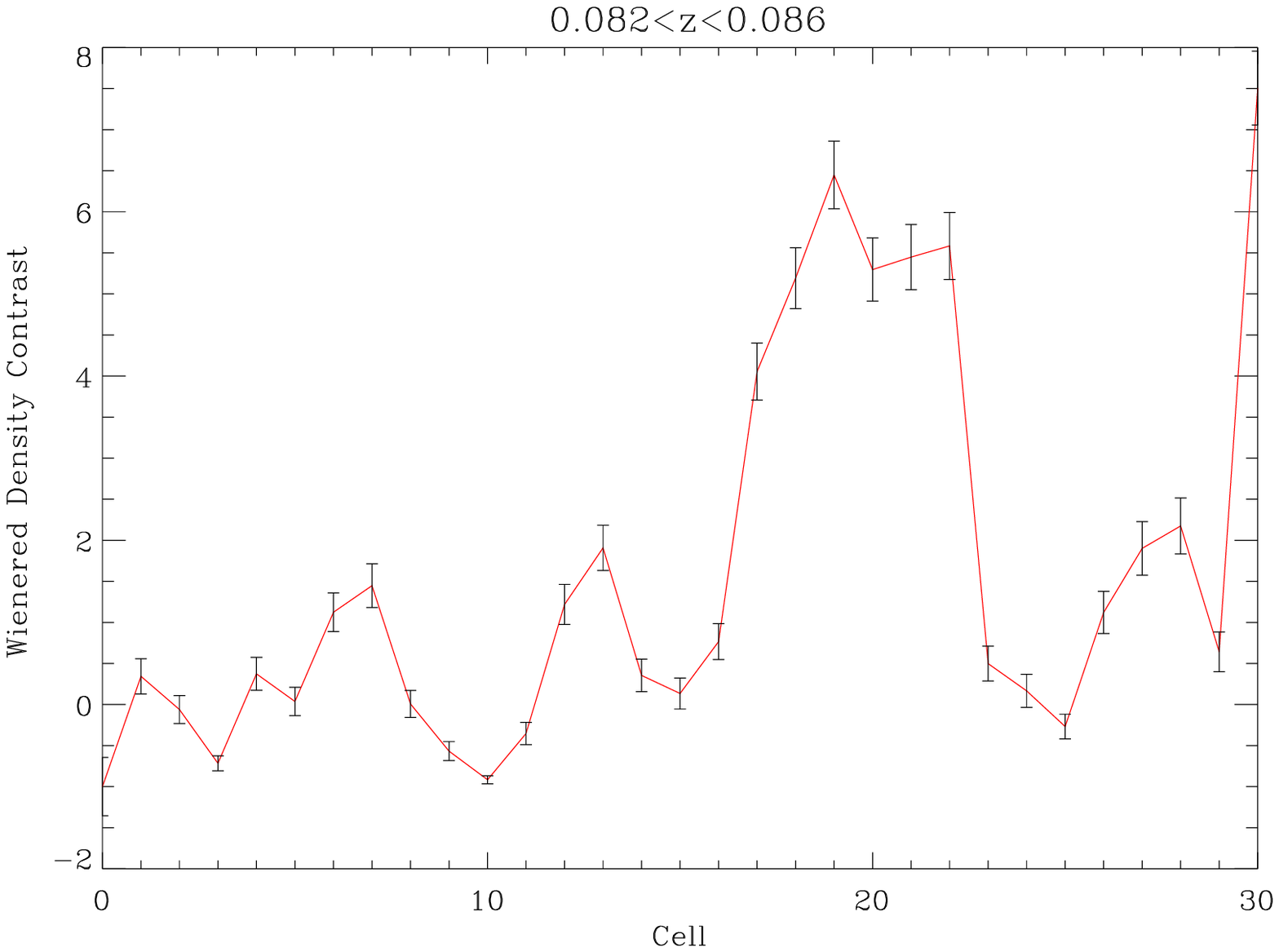,angle=0,width=0.25\textwidth, clip=} &
\psfig{figure=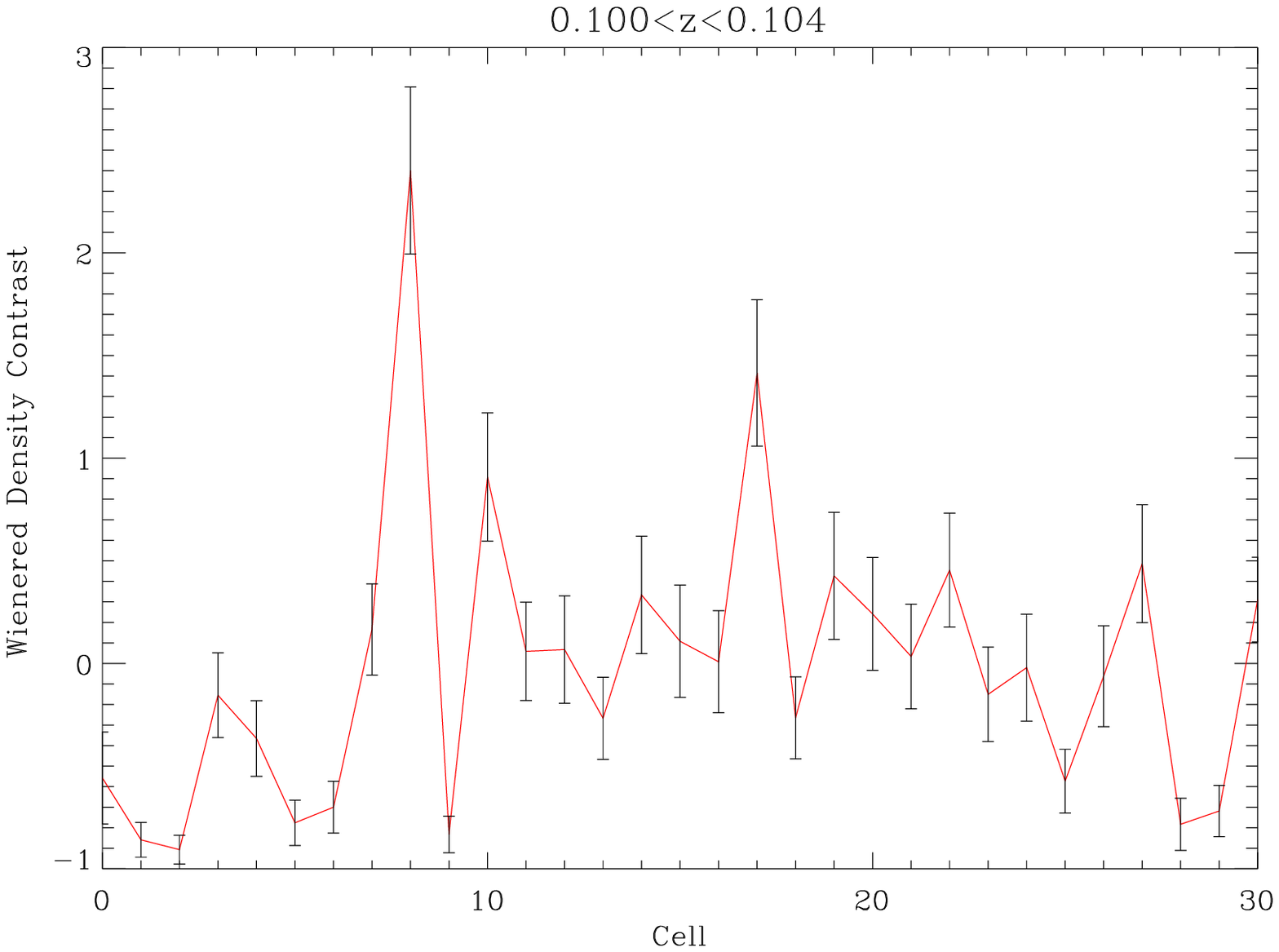,angle=0,width=0.25\textwidth, clip=} \\
\end{array}$
\caption[] {Same as in figure ~\ref{1res} but for the redshift
slices in NGP shown above.}
\label{2res}
\end{figure*}

We also plot the square root of the variance of the residual field
(equation~\ref{vares}), which defines the scatter around the mean
reconstructed field. We plot the residual fields corresponding to some
of the redshift slices shown in this paper.
(Figures~\ref{1res} and ~\ref{2res}). For better comparison, plots are made so that the cell number
increases with increasing $RA$.  If the volume of the cells used to
pixelise the survey was constant, we would expect the square root of
the variance $\Delta\delta$ to increase due to the increase in shot
noise (equation~\ref{res}).  However, since the pixelisation was
constructed to keep the shot noise per pixel approximately constant,
$\Delta\delta$ also stays constant ($\Delta\delta\approx0.23$ for both
NGP and SGP) but the average density contrast in each pixel decreases
with increasing redshift. This means that although the variance of the
residual in each cell is roughly equal, the relative variance
(represented by $\frac{\Delta\delta}{\delta}$) increases with 
increasing redshift. This increase is clearly evident in
Figures~\ref{1res} and ~\ref{2res}. Another conclusion that can be
drawn from the figures is that 
the bumps in the density field are due to real features not due to
error in the reconstruction, even at higher redshifts. 

\subsection{Reconstruction Using Non-linear Theory}

In order to increase the resolution of the density field maps, we
reduce the target cell width to $5 h^{-1}$ Mpc. A volume of a cubic
cell of side $5h^{-1}$ Mpc is roughly equal to a top-hat sphere of
radius of about $3 h^{-1}$ Mpc. The variance of the mass density field in
this sphere is $\sigma_{3}\approx1.7$ which corresponds to non-linear scales. To reconstruct the density field on
these scales, we require accurate descriptions of non-linear galaxy power
spectrum and the non-linear redshift space distortions. 

For the non-linear matter power spectrum $P_{\rm nl}^{R}(k)$, we adopt the
empirical fitting formula of Smith \etal (2003). This formula, derived
using the `halo model' for galaxy clustering, is more accurate than
the widely used Peacock \& Dodds (1996) fitting formula which is based
on the assumption of `stable clustering' of virialized halos.  We
note that for the scales concerned in this paper (up to $k\approx10 h
{\rm Mpc^{-1}}$), Smith \etal (2003) and Peacock \& Dodds (1996)
fitting formulae give very similar results. For simplicity we assume
linear, scale independent biasing in order to determine the galaxy power
spectrum from the mass power spectrum, where $b$ measures the ratio
between galaxy and mass distribution, 
\beqn
P_{\rm nl}^{R} = b^2P_{\rm nl}^{m}, 
\label{bias}
\eeqn
where $P_{\rm nl}^{R}$ is the galaxy and  $P_{\rm nl}^{m}$ is the matter
power spectrum.
We assume that $b=1.0$ for our analysis. While this value is in
agreement with the result obtained from 2dFGRS (Lahav \etal 2002,
Verde \etal 2002) for
scales of tens of Mpc, it does not hold true for the scales of $5 h^{-1}$ Mpc
on which different galaxy populations show different clustering patterns
(Norberg \etal 2002, Madgwick \etal 2002, Zehavi \etal 2002). More
realistic models exist where biasing is scale dependent (e.g. Seljak,
2000 and Peacock \& Smith, 2000) but since the Wiener Filtering method
is not sensitive to small errors in the prior parameters and the
reconstruction scales are not highly non-linear, the simple assumption of no
bias will still give accurate reconstructions. 

\begin{figure}
\psfig{figure=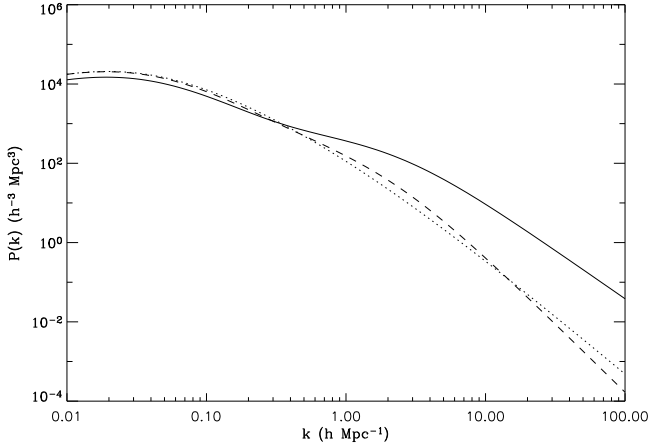,angle=0,width=0.5\textwidth,clip=}
\caption[] {Non-linear power spectra for z=0 and the concordance
model with $\sigma_{p}=506 {\rm km} s^{-1}$ in real space (solid
line), in redshift space from equation~\ref{nlpz} (dashed
line), both derived using the fitting 
formulae of Smith \etal (2003) and linear power spectra in
redshift space derived using linear theory and Kaiser's factor (dotted line).}
\label{pk} 
\end{figure}

The main effect of redshift distortions on non-linear scales is the
reduction of power as a result of radial smearing due to virialized
motions. The density profile in redshift space is then the convolution
of its real space counterpart with the peculiar velocity distribution
along the line of sight, leading to damping of power on small
scales. This effect is known to be reasonably well approximated by
treating the pairwise peculiar velocity field as Gaussian or better
still as an exponential in real space (superpositions of Gaussians),
with dispersion $\sigma_{p}$ (e.g. Peacock \& Dodds 1994, Ballinger
\etal 1996 \& Kang \etal 2002). Therefore the galaxy power spectrum in
redshift space is written as 
\beqn
P_{\rm nl}^{S}(k,\mu)= P_{\rm nl}^{R}(k,\mu)(1+\beta\mu^2)^2D(k\sigma_p\mu)\;,
\label{nlps}
\eeqn where $\mu$ is the cosine of the wave vector to the line of sight,
$\sigma_{p}$ has the unit of $h^{-1}$ Mpc and the damping function
in $k$-space is a Loretzian: 
\beqn
D(k\sigma_p\mu)=\frac{1}{1+(k^2\sigma_p^2\mu^2)/2}.  
\eeqn 
Integrating
equation~\ref{nlps} over $\mu$, we obtain the direction-averaged power
spectrum in redshift space: 
\beqn
 \frac{P_{\rm nl}^{S}(k)}{P_{\rm nl}^{R}(k)}
&=&\frac{4(\sigma_p^2k^2-\beta)\beta}{\sigma_p^4k^4}+\frac{2\beta^2}{3\sigma_p^2k^2}\nonumber\\
&+&\frac{\sqrt{2}(k^2\sigma_p^2-2\beta)^2\mbox{arctan}(k\sigma_p/\sqrt{2})}{k^5\sigma_p^5}.
\label{nlpz}
\eeqn 
For the non-linear reconstruction, we use
equation~\ref{nlpz} instead of equation~\ref{linpz} when deriving the
correlation function in redshift space. Figure~\ref{pk} shows how
the non-linear power spectrum is damped in redshift space (dashed
line) and compared to the linear power spectrum (dotted line).  In
this plot and throughout this paper we adopt the 
$\sigma_{p}$ value derived by Hawkins \etal (2002),
$\sigma_{p}=506\pm52 {\rm km} s^{-1}$. Interestingly, by coincidence, the
non-linear and linear power spectra look very similar in redshift
space. So, if we had used the linear power spectrum instead of
its non-linear counterpart, we still would have obtained physically
accurate reconstructions of the density field in redshift space.

The optimal density field in real space is calculated using
equation~\ref{opt}. The cross-correlation matrix in
equation~\ref{appopt} can now be approximated as
\begin{equation}
\langle {\bf s}({\bf r}) {\bf d}({\bf s})\rangle = \xi(r,\mu)
(1+\beta\mu^2)\sqrt{D(k\sigma_p\mu)}\;.
\end{equation}
Again, integrating the equation above over $\mu$, the direction averaged
cross-correlation matrix of the density field in real space and the
density field in redshift space can be written as
\beqn
\frac{\langle {\bf s}({\bf r}) {\bf d}({\bf s})\rangle}{\langle {\bf s}({\bf r}) {\bf s}({\bf 
r})\rangle}
&=&\frac{1}{2k^2\sigma_p^2}\mbox{ln}(k^2\sigma_p^2(1+\sqrt{1+1/k^2\sigma_p^2}))\nonumber\\
&+&\frac{\beta}{k^2\sigma_p^2}\sqrt{1+k^2\sigma_p^2}+\frac{\beta}{k^3\sigma_p^3}\mbox{arcsinh}(k^2\sigma_p^2).
\eeqn

At the end of this paper, we show some examples of the non-linear
reconstructions (Figures~\ref{00sgp},~~\ref{5mpc10sgp}, \ref{5mpc2ngp}
and~\ref{5mpc6ngp}). As can be seen from these plots the resolution of
the reconstructions improves radically, down to the scale of large
clusters.  Comparing Figure~\ref{14ngp} and Figure~\ref{5mpc6ngp}
where the redshift ranges of the maps are similar, we
conclude that 10 $h^{-1}$ Mpc and 5 $h^{-1}$ Mpc resolutions give
consistent reconstructions.

Due to the very large number of cells, we
reconstruct
four separate density fields for redshift ranges $0.035\lsim$z$\lsim0.05$ 
and
$0.09\lsim$z$\lsim0.11$ in SGP, $0.035\lsim$z$\lsim0.05$ and 
$0.09\lsim$z$\lsim0.12$ in NGP. There are
46 redshift slices in total. The number of cells for the
latter reconstructions are in the order of 10 000.

To investigate the effects of
using different non-linear redshift distortion approximations, we also
reconstruct one field without the damping function but just collapsing
the fingers of god. Although including the damping function results
in a more physically accurate reconstruction of the density field in
real space, it still is not enough to account for the elongation of
the richest clusters along the line of sight. Collapsing the fingers
of god as well as including the damping function underestimates power
resulting in noisy reconstruction of the density field. We choose to
use only the damping function for the non-linear scales obtaining 
stable results for the density field reconstruction.

The theory of gravitational instability states that as the dynamics
evolve away from the linear regime, the initial field deviates from
Gaussianity and skewness develops. The Wiener Filter, in the form
presented here, only minimises the
variance and it ignores the higher moments that describe the skewness of
the underlying distribution.
However, since the scales of reconstructions presented in this section 
are not highly non-linear and the signal-to-noise ratio is 
high, the assumptions involved in this analysis are not severely
violated for the non-linear reconstruction of the density field in
redshift space. The real space reconstructions are more sensitive to the
choice of cell size and the power spectrum since the Wiener Filter is
used not only for noise suppression but also for transformation from
redshift space. Therefore, reconstructions in redshift space are
more reliable on these non-linear scales than those in real space. 

A different approach to the non-linear density field reconstruction
presented in this paper is to apply the 
Wiener Filter to the reconstruction of the logarithm of the density
field as there is good evidence that the statistical properties of the
perturbation field in the quasi-linear regime is well approximated by
a log-normal distribution. A detailed analysis of the application of
the Wiener Filtering technique to log-normal fields is given in Sheth
(1995).    

\section{Mapping the Large Scale Structure of the 2\lowercase{d}FGRS}

One of the main goals of this paper is to use the reconstructed
density field to identify the major superclusters and voids in the
2dFGRS. We define superclusters (voids) as
regions of large overdensity (underdensity) which is above (below) a
certain threshold. This approach has been used successfully by several
authors (e.g. Einasto \etal 2002 \& 2003, Kolokotronis, Basilakos \&
Plionis 2002, Plionis \& Basilakos 2002, Saunders \etal 1991). 

\subsection{The Superclusters}
\begin{figure}
\psfig{figure=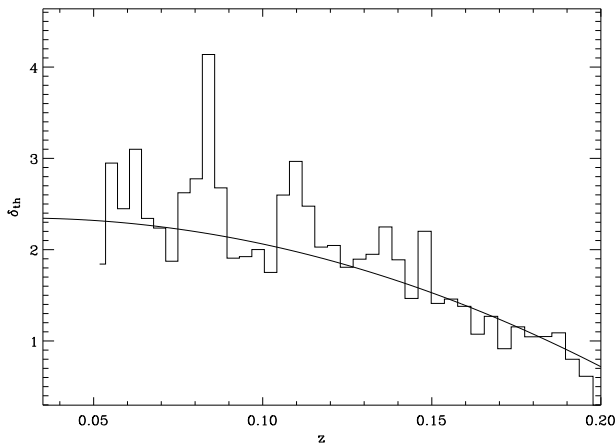, width=0.5\textwidth, angle=0}
\caption[] {The density threshold $\delta_{th}$ as a function of
redshift $z$ and the best fit model used to select the superclusters
in Table 1.}
\label{thresh}
\end{figure}

In
order to find the superclusters listed in Table~\ref{tab:sc}, we
define the density contrast threshold, $\delta_{th}$, as distance
dependent. We use a varying density threshold for two reasons that are
related to the way the density field is reconstructed. Firstly, the
adaptive gridding we use implies that the cells get bigger with
increasing redshift. This means that the density contrast in each cell
decreases. The second effect that decreases the density contrast
arises due to the Wiener Filter signal tending to zero towards the edges of the
survey. Therefore, for each redshift slice we find the mean and the
standard deviation of the density field (averaged over 113 cells for
NGP and 223 for SGP), then we calculate $\delta_{th}$ as twice the
standard deviation of the field added to its mean, averaged over SGP
and NGP for each redshift bin (see Figure~\ref{thresh}). In order to
account for 
the clustering effects, we fit a smooth curve to $\delta_{th}$, using
$\chi^2$ minimisation. The best fit, also shown Figure~\ref{thresh}, is a
quadratic equation with $\chi^2$/(number of degrees of freedom) =
1.6. We use this fit when selecting the overdensities. We note that
choosing  smoothed or unsmoothed density threshold does not
change the selection of the superclusters.

The list of superclusters in SGP and NGP region are given in
Table~\ref{tab:sc}. This table is structured as follows: column 1 is
the identification, columns 2 and 3 are the minimum and the maximum
redshift, columns 4 and 5 are the minimum and maximum $RA$, columns 6
and 7 are the minimum and maximum $Dec$ over which the density contours above
$\delta_{th}$ extend. In column 8, we show the number of Durham
groups with more than 5 members that the supercluster contains. In column
9, we show the number of Abell, APM and EDCC clusters studied by De
Propris \etal (2002) the supercluster has. In the last column, we show
the total number of groups and clusters. Note that most of the Abell
clusters are counted in the Durham group catalogue. We observe that the
rich groups (groups with more than 9 members) almost always reside in
superclusters whereas poorer groups are more dispersed. We note that
the superclusters that contain Abell clusters are on average richer
than superclusters that do not contain Abell clusters, in agreement
with Einasto \etal 2003. However, we also note that the number of
Abell clusters in a rich supercluster can be equal to the number of
Abell clusters in a poorer supercluster, whereas the number of Durham
groups increase as the overdensity increases. Thus, we conclude that
the Durham groups are in general better representatives of the
underlying density distribution of 2dFGRS than Abell clusters.

The superclusters SCSGP03,
SCSGP04 and SCSGP05 can be seen in Figure~\ref{02sgp}. SCSGP04 is part
the rich Pisces-Cetus Supercluster which was first described by Tully
(1987). SCSGP01, SCSGP02, SCSGP03 and
SCSGP04 are all filamentary structures connected to each other,
forming a multi branching system. SCSGP05 seems to be a more isolated
system, possibly connected to SCSGP06 and SCSGP07. SCSGP06
(figure~\ref{06sgp}) is the upper part of the gigantic Horoglium Reticulum
Supercluster. SCSGP07 (Figure~\ref{06sgp}) is the extended part of the
Leo-Coma Supercluster. The richest supercluster in
the SGP region is SCSGP16 which can be seen in the middle of the plots
in Figure~\ref{5mpc10sgp}. Also shown, in the same figure, is SCSGP15,
one of the richest superclusters in SGP and the edge of SCSGP17. In
fact, as evident from Figure~\ref{5mpc10sgp}, SCSGP14,
SCSGP15, SCSGP16 and SCSGP17 are branches of one big
filamentary structure. In Figure~\ref{5mpc2ngp}, we see SCNGP01, this
structure is part of the upper edge of the Shapley Supercluster. The richest
supercluster in
NGP is SCNGP06, shown in Figure~\ref{09ngp}. Around this, in the
neighbouring redshift slices, there are
two rich filamentary superclusters, SCNGP06, SCNGP08 (figure~\ref{14ngp}). 
SCNGP07 seems to
be the node point of these filamentary structures. 

Table 1 shows only the major overdensities in the survey. These tend
to be filamentary structures that are mostly connected to each
other.

\subsection{The Voids}
For the catalogue of the largest voids in Table~\ref{tab:sgpvoid} and 
Table~\ref{tab:ngpvoid}, we
only consider regions with 80 percent completeness or more and go up
to a redshift of 0.15. Following the previous studies (e.g. El-Ad \&
Piran 2000, Benson \etal 2003 \& Sheth \etal 2003), our chosen
underdensity threshold is $\delta_{th}=-0.85$. The lists of galaxies
and their properties that are in these voids and in the other smaller
voids that are not in Tables~\ref{tab:sgpvoid} and~\ref{tab:ngpvoid}  are 
given in url:
http://www.ast.cam.ac.uk/$\sim$pirin.
The tables of voids is structured in a similar way as the table of
superclusters, only we give separate tables for NGP and SGP voids.
The most intriguing structures seen are the voids VSGP01, VSGP2,
VSGP03, VSGP4 and VSGP05. These voids, clearly visible in the radial number 
density function
of 2dFGRS, are actually part of a big superhole
broken up by the low value of the void density threshold. This
superhole has been observed and investigated by several authors
(e.g. Cross \etal 2002, De Propis  \etal 2002, Norberg \etal 2002, Frith \etal 2003). The local NGP region also
has excess underdensity (VNGP01,VNGP02, VNGP03, VNGP04, VNGP05, VNGP06
and VNGP07, again part of a big superhole) but the voids in this area
are not as big or as empty as the voids in local SGP. In fact, by
combining the results from 2 Micron All Sky Survey, Las Campanas
Survey and 2dFGRS, Frith \etal (2003) conclude that these underdensities
suggest that there is a contiguous void stretching from north to south.
If such a void does exist then it is unexpectedly large for our
present understanding of large scale structure, where on large enough
scales the Universe is isotropic and homogeneous.

\section{Conclusions}

In this paper we use the Wiener Filtering technique to reconstruct
the density field of the 2dF galaxy redshift survey. We pixelise the
survey into igloo cells bounded by $RA$, $Dec$ and redshift. The cell
size varies in order to keep the number of galaxies per cell roughly
constant and is approximately 10 $h^{-1}$ for Mpc high and 5 $h^{-1}$ Mpc for
the low resolution maps at the median redshift
of the survey. Assuming a prior based on parameters $\omegam = 0.3$,
$\omegal = 0.7$, $\beta = 0.49$, $\sigma_8 = 0.8$ and $\Gamma = 0.2$,
we find that the reconstructed density field clearly picks out the
groups catalogue built by Eke \etal (2003) and Abell, APM and EDCC
clusters investigated by De Propris \etal (2002). We also reconstruct
four separate density fields with different redshift ranges for a smaller
cell size of 5 $h^{-1}$ Mpc at
the median redshift. For these reconstructions, we assume a non-linear
power spectrum fit developed by Smith \etal (2003) and linear biasing. The resolution of
the density field improves dramatically, down to the size of big
clusters. The derived high resolution density fields is in agreement
with the lower resolution versions.

We use the reconstructed fields to identify the major superclusters
and voids in SGP and NGP. We find that the richest superclusters are
filamentary and multi branching, in agreement with Einasto \etal
2002. We also find that the rich clusters always reside in
superclusters whereas poor clusters are more dispersed. We present the
major superclusters in 2dFGRS in Table 1. We pick out two very rich
superclusters, one in SGP and one in NGP. We also identify voids as underdensities
that are below $\delta\approx-0.85$ and that lie in regions with more
that 80$\%$ completeness. These underdensities are presented
in Table 2 (SGP) and Table 3 (NGP). We pick out two big voids, one
in SGP and one in NGP. Unfortunately, we cannot
measure the sizes and masses of the large structures we observe in
2dF as most of these structures continue beyond the boundaries of the
survey. 

The detailed maps and lists of all of the reconstructed density
fields, plots of the 
residual fields and the lists of the galaxies that are in underdense
regions  can be found on WWW at 
http://www.ast.cam.ac.uk/$\sim$pirin.

One of the main aims of this paper is to identify the large scale
structure in 2dFGRS. The Wiener Filtering technique provides a
rigorous methodology for variable smoothing and noise suppression. As
such, a natural continuation of this work is using the Wiener
filtering method in conjunction with other methods to further
investigate the geometry and the topology of the supercluster-void
network. Sheth \etal (2003) developed a powerful surface modelling
scheme, SURFGEN, in order to calculate the Minkowski Functionals of the
surface generated from the density field. The four Minkowski
functionals -- the area, the volume, the extrinsic curvature and the
genus -- contain information about the geometry, connectivity and
topology of the surface (cf. Mecke, Buchert \& Wagner 1994 and Sheth
\etal 2003) and thus they will provide a detailed morphological
analysis of the superclusters and voids in 2dFGRS.

Although not applicable to 2dFGRS, the Wiener reconstruction technique
is well suited to recovering the velocity
fields from peculiar velocity catalogues. Comparisons of galaxy
density and velocity fields allow  direct estimations of the 
cosmological parameters such as the bias 
parameter and the mean mass density. These comparisons will be
possible with the upcoming 6dF Galaxy Survey
(http://www..mso.anu.edu.au/6dFGS)
which will measure the redshifts of 170,000 galaxies and the peculiar
velocities of 15,000 galaxies by June 2005. Compared to 2dFGRS, the
6dF survey has a much higher sky-coverage (the
entire southern sky down to $|b|>10^\circ$). This wide survey area
would allow a full hemispheric Wiener reconstruction
of large scale structure so that the sizes and masses of the
superclusters and voids can be determined.

\section*{ACKNOWLEDGEMENTS}

The 2dF Galaxy Redshift Survey was made possible through the dedicated
efforts of the staff at Anglo-Australian Observatory, both in
crediting the two-degree field instrument and supporting it on the
telescope. PE acknowledges support from the Middle East Technical
University, Ankara, Turkey, the Turkish Higher Education Council and
Overseas Research Trust. This research was conducted utilising the
super computer COSMOS at DAMTP, Cambridge. 


\begin{figure*}
\caption[] {Reconstructions of the 2dFGRS SGP region for the redshift
range: $0.047 \leq z \leq 0.049$ for $5 h^{-1}$
Mpc target cell size. The contours are spaced
at $\Delta\delta = 0.5$ with solid (dashed) lines denoting positive
(negative) contours; the heavy solid contours correspond to $\delta =
0$. The red dots denote the galaxies with redshifts in the plotted
range. a) Redshift space density field weighted by the selection function
and the angular mask. b) Same as in a) but smoothed by a Wiener Filter
c) Same as in b) but corrected for the redshift distortion. The
overdensity centred on: 1) $RA\approx-23.5$, $Dec\approx-30.0$ is SCSGP03
(see Table 1); 2) $RA\approx0.0$, $Dec\approx-30.0$ is SCSGP04; 3)
$RA\approx36.0$, $Dec\approx-29.3$ is SCSGP05. The
underdensity centred on $RA\approx-20.5$, $Dec\approx-30.0$ is
VSGP01; 2) $RA\approx-8.5$, $Dec\approx-29.3$ is VSGP02; 3)
$RA\approx18.0$, $Dec\approx-28.5$ is VSGP04; 4) $RA\approx32.2$, 
$Dec\approx-29.5$ is VSGP05.
(see Table 2).}
\label{00sgp}
\end{figure*}

\begin{figure*}
\caption[] {Reconstructions of the 2dFGRS SGP region for the redshift
range: $0.057 \leq z \leq 0.061$ for $10 h^{-1}$
Mpc target cell size. Same as in figure~\ref{00sgp}.The
overdensity centred on: 1) $RA\approx-23.5$, $Dec\approx-30.0$ is SCSGP03
(see Table 1); 2) $RA\approx0.0$, $Dec\approx-30.0$ is SCSGP04, this
overdensity is part of the Pisces-Cetus Supercluster; 3)
$RA\approx36.0$, $Dec\approx-29.3$ is SCSGP05. The
underdensity centred on $RA\approx-10.0$, $Dec\approx-30.0$ is VSGP12
(see Table 2).}
\label{02sgp}
\end{figure*}

\begin{figure*}
\caption[] {Reconstructions of the 2dFGRS SGP region for the redshift
range: $0.068 \leq z \leq 0.071$ for $10 h^{-1}$
Mpc target cell size. Same as in figure~\ref{00sgp}.The
overdensity centred on: 1) $RA\approx39.0$, $Dec\approx-34.5$ is SCSGP07
(see Table 1) and is part of Leo-Coma Supercluster; 2) $RA\approx0.0$,
$Dec\approx-30.0$ is SCSGP06 and is part of Horoglium Reticulum 
Supercluster;
(see Table 1).}
\label{06sgp}
\end{figure*}

\begin{figure*}
\caption[] {Reconstructions of the 2dFGRS SGP region for the redshift
range: $0.107 \leq z \leq 0.108$ for $5 h^{-1}$
Mpc target cell size. Same as in
figure~\ref{00sgp}.The
overdensity centred on: 1) $RA\approx1.7$, $Dec\approx-31.0$ is SCSGP16
(see Table 1); 2) $RA\approx36.3$, $Dec\approx-30.0$ is SCSGP15. 3)
$RA\approx-15.0$, $Dec\approx-30.0$ is SCSGP17.The
underdensity centred on: 1) $RA\approx-25.0$, $Dec\approx-35.2$ is VSGP25
(see Table 2); 2) $RA\approx11.3$, $Dec\approx-24.5$ is VSGP22; 3) 
$RA\approx48.0$, $Dec\approx-30.5$ is VSGP20.}
\label{5mpc10sgp}
\end{figure*}

\begin{figure*}
\caption[] {Reconstructions of the 2dFGRS NGP region for the redshift
range: $0.039 \leq z \leq 0.041$  for $5 h^{-1}$
Mpc target cell size. Same as in figure~\ref{00sgp}.The
overdensity centred on $RA\approx153.0$, $Dec\approx-4.0$ is SCNGP01
and is part of the Shapley Supercluster (see Table 1). The
underdensities are VNGP01, VNGP02, VNGP03, VNGP04, VNGP05, VNGP06 and
VNGP07 (see Table 3)}
\label{5mpc2ngp}
\end{figure*}

\begin{figure*}
\caption[] {Reconstructions of the 2dFGRS NGP region for the redshift
range: $0.082 \leq z \leq 0.086$  for $10 h^{-1}$
Mpc target cell size. Same as in figure~\ref{00sgp}.The
overdensity centred on $RA\approx194.0$, $Dec\approx-2.5$ is SCNGP06
(see Table 1).}
\label{09ngp}
\end{figure*}

\begin{figure*}
\caption[] {Reconstructions of the 2dFGRS NGP region for the redshift
range: $0.100 \leq z \leq 0.104$  for $10 h^{-1}$
Mpc target cell size. Same as in figure~\ref{00sgp}.The
overdensity centred on: 1) $RA\approx170.0$, $Dec\approx-1.0$ is SCNGP08
(see Table 1). The
underdensity centred on: 1) $RA\approx150.0$, $Dec\approx-1.5$ is VNGP18
(see Table 3); 2) $RA\approx192.5$, $Dec\approx0.5$ is VNGP19; 2) 
$RA\approx209.0$, $Dec\approx-1.5$ is VNGP17.}
\label{14ngp}
\end{figure*}

\begin{figure*}
\caption[] {Reconstructions of the 2dFGRS NGP region for the redshift
range: $0.103 \leq z \leq 0.108$  for $5 h^{-1}$
Mpc target cell size. Same as in figure~\ref{00sgp}.}
\label{5mpc6ngp}
\end{figure*}

\scriptsize
\vskip-2.5cm
\begin{table*}
\begin{center}
\caption{The list of superclusters}
\begin{tabular}{@{}cccccccccc}
\\ \hline \\ $No$ & $z_{min}$ & $z_{max}$ & $RA_{min}$ & $RA_{max}$ &
$Dec_{min}$ & $Dec_{max}$ & $N_{gr} \geq 9$ & $N_{clus}$ & $N_{total}$
\\ & & & (1950) deg & (1950) deg & (1950) deg & (1950) deg & & \\ \\
\hline \\

SCSGP01 & 0.048 & 0.054 & $-$11.5 & $-$3.2 & $-$37.5 & $-$25.5 & 23 &
9 & 27 \\

SCSGP02 & 0.054 & 0.057 & $-$22.3 & 0.0 & $-$35.5 & $-$33.0 & 9 & 1 &
9 \\

SCSGP03 & 0.057 & 0.064 & $-$29.4 & $-$17.5 & $-$37.5 & $-$22.5 & 23&
8 & 26 \\

SCSGP04 & 0.057 & 0.068 & $-$10.0 & +10.0 & $-$35.5 & $-$24.0 & 34 &
14 & 37 \\

SCSGP05 & 0.054 & 0.064 & 32.5 & 39.0 & $-$33.0 & $-$25.5 & 14 & 7 &
18 \\

SCSGP06 & 0.064 & 0.082 & 45.5 & 55.0 & $-$35.0 & $-$24.0 & 26 & 7 &
31 \\

SCSGP07 & 0.064 & 0.071 & 18.5 & 40.0 & $-$35.0 & $-$31.0 & 10 & 4 &
11\\

SCSGP08 & 0.068 & 0.075 & $-$38.5 & $-$33.0 & $-$35.0 & $-$22.5 & 9 &
3 & 11 \\

SCSGP09 & 0.075 & 0.082 & $-$27.0 & $-$17.5 & $-$37.5 & $-$22.5 & 20 &
9 & 24 \\

SCSGP10 & 0.082 & 0.093 & $-$17.5 & $-$9.5 & $-$36.0 & $-$29.0 & 18 &
9 & 21 \\

SCSGP11 & 0.086 & 0.097 & $-$33.0 & $-$27.0 & $-$34.5 & $-$24.0 & 14 &
4 & 15 \\

SCSGP12 & 0.093 & 0.097 & $-$22.0 & $-$17.4 & $-$36.0 & $-$34.8 & 4 &
1 & 5 \\

SCSGP13 & 0.093 & 0.097 & 0.0 & 3.8 & $-$36.0 & $-$34.8 & 3 & 0 & 3 \\

SCSGP14 & 0.093 & 0.104 & 16.0 & 24.6 & $-$33.0 & $-$28.2 & 8 & 1 & 8
\\

SCSGP15 & 0.097 & 0.115 & 28.0 & 44.6 & $-$35.2 & $-$24.6 & 34 & 18 &
41 \\

SCSGP16 & 0.100 & 0.119 & 1.5 & 21.9 & $-$35.2 & $-$26.5& 93 & 20 & 98
\\

SCSGP17 & 0.100 & 0.108 & $-$27.1 & $-$3.0 & $-$35.2 & $-$26.5& 38 & 7
& 43 \\

SCSGP18 & 0.150 & 0.166 & 1.5 & 31.8 & $-$33.0 & $-$26.5 & 18 & 6 & 22
\\

SCSGP19 & 0.162 & 0.177 & 22.8 & 34.5 & $-$35.2 & $-$24.6 & 13 & 5 &
15 \\

SCSGP20 & 0.181 & 0.202 & $-$17.5 & $-$3.15 & $-$35.5 & $-$24.6 & 13 &
4 & 14 \\

\\

SCNGP01 & 0.035 & 0.068 & 147.5 & 174.5 & $-$6.5 & 2.5 & 59 & 6 & 61\\

SCNGP02 & 0.048 & 0.061 & 210.0 & 218.5 & $-$5.0 & 0.0 & 19 & 0 & 19\\

SCNGP03 & 0.068 & 0.075 & 150.0 & 155.0 & $-$1.0 & 2.5 & 6 & 0 & 6 \\

SCNGP04 & 0.068 & 0.075 & 158.0 & 166.5 & $-$1.0 & 2.5 & 8 & 1 & 9 \\

SCNGP05 & 0.071 & 0.082 & 171.0 & 184.0 & $-$3.5 & 2. 5 & 27 & 6 & 28
\\

SCNGP06 & 0.079 & 0.094 & 185.0 & 202.5 & $-$7.5 & 2.5 & 77 & 7 & 79
\\

SCNGP07 & 0.086 & 0.101 & 147.0 & 181.5 & $-$5.5 & 2.5 & 31 & 4 & 34
\\

SCNGP08 & 0.090 & 0.131 & 165.0 & 185.0 & $-$6.3 & 2.5 & 51 & 10 & 57
\\

SCNGP09 & 0.103 & 0.114 & 158.5 & 163.0 & $-$1.3 & 2.5 & 8 & 1 & 9 \\

SCNGP10 & 0.103 & 0.118 & 197.0 & 203.0 & $-$3.75 & 1.25 & 22 & 0 & 22
\\

SCNGP11 & 0.123 & 0.131 & 147.0 & 173.0 & 0.0 & 2.5 & 3 & 2 & 3 \\

SCNGP12 & 0.119 & 0.142 & 210.0 & 219.0 & $-$4.0 & 1.3& 19 & 2 & 20 \\

SCNGP13 & 0.131 & 0.142 & 173.0 & 179.0 & $-$6.25 & $-$1.3 & 5 & 2 & 6
\\

SCNGP14 & 0.131 & 0.142 & 185.0 & 193.0 & $-$6.25 & 1.3 & 7 & 0 & 7 \\

SCNGP15 & 0.131 & 0.138 & 155.0 & 160.0 & $-$6.3 & 1.3 & 9 & 1 & 9 \\

SCNGP16 & 0.142 & 0.146 & 166.0 & 174.0 & $-$0.3 & 1.3 & 5 & 0 & 5 \\

SCNGP17 & 0.142 & 0.146 & 194.0 & 199.0 & $-$2.5 & 1.3 & 1& 0 & 1 \\

SCNGP18 & 0.146 & 0.150 & 204.0 & 210.0 & $-$2.5 & 1.3 & 4& 0 & 4 \\

SCNGP19 & 0.173 & 0.177 & 181.0 & 184.0 & $-$1.0 & $-$4.0 & 3& 0 & 3
\\

SCNGP20 & 0.181 & 0.185 & 185.5 & 193.0 & $-$1.0 & $-$4.0 & 1& 0 & 1
\\

\\ \hline
\label{tab:sc}
\end{tabular}
\end{center}.
\end{table*}

\scriptsize
\vskip-2.5cm
\begin{table*}
\begin{center}
\caption{The list of voids in SGP}
\begin{tabular}{@{}cccccccccc}
\\ \hline \\ $No$ & $z_{min}$ & $z_{max}$ & $RA_{min}$ & $RA_{max}$ &
$Dec_{min}$ & $Dec_{max}$ \\ & & & (1950) deg & (1950) deg & (1950)
deg & (1950) deg \\ \\

VSGP01 & 0.035 & 0.051 & $-$27.5 & $-$17.3 & $-$37.5 & $-$22.5 \\

VSGP02 & 0.035 & 0.051 & $-$17.3 & 0.0 & $-$33.0 & $-$25.5 \\

VSGP03 & 0.035 & 0.048 & 0.9 & 11.5 & $-$34.5 & $-$28.5 \\

VSGP04 & 0.035 & 0.051 & 12.4 & 23.6 & $-$30.5 & $-$26.0 \\

VSGP05 & 0.035 & 0.054 & 23.6 & 40.8 & $-$33.5 & $-$26.0 \\

VSGP06 & 0.035 & 0.042 & 32.8 & 37.3 & $-$33.5 & $-$28.5 \\

VSGP07 & 0.035 & 0.051 & 39.5 & 46.4 & $-$31.5 & $-$28.5 \\

VSGP08 & 0.035 & 0.042 & 46.4 & 51.0 & $-$31.5 & $-$28.5 \\

VSGP09 & 0.039 & 0.044 & $-$35.5 &$-$30.0 & $-$34.5 & $-$28.5 \\

VSGP10 & 0.039 & 0.041 & 10.0 & 17.5 & $-$33.5 & $-$29.5 \\

VSGP11 & 0.041 & 0.044 & 15.5 & 22.5 & $-$33.5 & $-$28.5 \\

VSGP12 & 0.057 & 0.061 & $-$12.8 & $-$8.2 & $-$31.5 & $-$28.5 \\

VSGP13 & 0.082 & 0.086 & 5.4 & 8.0 & $-$31.5 & $-$28.5 \\

VSGP15 & 0.082 & 0.089 & 41.8 & 49.3 & $-$31.5 & $-$28.5 \\

VSGP16 & 0.086 & 0.089 & 12.5 & 16.3 & $-$31.5 & $-$28.5 \\

VSGP17 & 0.086 & 0.089 & 31.0 & 33.5 & $-$32.5 & $-$28.5 \\

VSGP18 & 0.089 & 0.097 & $-$8.2 & $-$5.95 & $-$30.5 & $-$29.5 \\

VSGP19 & 0.089 & 0.093 & 42.7 & 45.5 & $-$30.5 & $-$29.5 \\

VSGP20 & 0.093 & 0.104 & 46.4 & 52.5 & $-$32.5 & $-$28.2 \\

VSGP21 & 0.093 & 0.100 & 28.2 & 32.8 & $-$31.5 & $-$28.5 \\

VSGP22 & 0.097 & 0.104 & 7.7 & 12.5 & $-$32.0 & $-$28.5 \\

VSGP23 & 0.097 & 0.100 & $-$12.8 & $-$11.5 & $-$32.0 & $-$28.5 \\

VSGP24 & 0.100 & 0.108 & $-$31.8 & $-$25.3 & $-$28.5 & $-$25.5 \\

VSGP25 & 0.100 & 0.112 & $-$25.3 & $-$21.9 & $-$34.5 & $-$33.0 \\

VSGP26 & 0.112 & 0.119 & $-$7.0 & $-$3.7 & $-$31.5 & $-$28.5 \\

VSGP27 & 0.112 & 0.115 & 1.5 & 5.4 & $-$30.0 & $-$27.0 \\

VSGP28 & 0.112 & 0.115 & 25.9 & 28.2 & $-$30.0 & $-$27.0 \\

VSGP29 & 0.112 & 0.115 & 41.0 & 46.4 & $-$30.0 & $-$27.0 \\

VSGP30 & 0.115 & 0.119 & 26.2 & 30.5 & $-$34.5 & $-$31.5 \\

VSGP31 & 0.119 & 0.123 & $-$35.0 & $-$30.8 & $-$28.5 & $-$25.5 \\

VSGP32 & 0.119 & 0.123 & $-$27.0 & $-$25.5 & $-$28.5 & $-$26.5 \\

VSGP33 & 0.119 & 0.123 & $-$12.8 & $-$10.0 & $-$28.5 & $-$25.5 \\

VSGP34 & 0.119 & 0.123 & 41.5 & 49.3 & $-$38.8 & $-$31.8 \\

VSGP35 & 0.119 & 0.127 & 11.2 & 16.5 & $-$28.2 & $-$27.5 \\

VSGP36 & 0.123 & 0.131 & $-$26.4 & $-$22.4 & $-$31.5 & $-$28.5 \\

VSGP37 & 0.123 & 0.127 & $-$12.8 & $-$8.2 & $-$30.0 & $-$27.0 \\

VSGP38 & 0.131 & 0.142 & $-$26.4 & $-$19.5 & $-$34.5 & $-$31.5 \\

VSGP39 & 0.131 & 0.138 & 32.3 & 37.5 & $-$35.2 & $-$24.8 \\

VSGP40 & 0.138 & 0.142 & 30.3 & 33.5 & $-$33.5 & $-$29.0 \\
\\

\hline
\label{tab:sgpvoid}
\end{tabular}
\end{center}.
\end{table*}

\scriptsize
\vskip-2.5cm
\begin{table*}
\begin{center}
\caption{The list of voids in NGP}
\begin{tabular}{@{}cccccccccc}
\\ \hline \\ $No$ & $z_{min}$ & $z_{max}$ & $RA_{min}$ & $RA_{max}$ &
$Dec_{min}$ & $Dec_{max}$ \\ & & & (1950) deg & (1950) deg & (1950)
deg & (1950) deg \\ \\

VNGP01 & 0.035 & 0.054 & 147.0 & 160.0 & $-$4.0 & 1.5 \\

VNGP02 & 0.035 & 0.041 & 166.0 & 171.0 & $-$1.0 & 2.5 \\

VNGP03 & 0.035 & 0.046 & 171.0 & 181.2 & $-$1.5 & 2.5 \\

VNGP04 & 0.035 & 0.051 & 183.0 & 189.0 & $-$1.5 & 2.5 \\

VNGP05 & 0.035 & 0.046 & 189.0 & 197.0 & $-$1.5 & 2.0 \\

VNGP06 & 0.035 & 0.046 & 202.2 & 206.0 & $-$1.5 & 2.0 \\

VNGP07 & 0.037 & 0.046 & 204.2 & 215.6 & $-$2.0 & 0.0 \\

VNGP08 & 0.041 & 0.051 & 163.8 & 169.8 & $-$4.5 & 2.5 \\

VNGP09 & 0.049 & 0.061 & 177.0 & 196.4 & $-$3.5 & 2.5 \\

VNGP10 & 0.054 & 0.057 & 165.5 & 173.0 & $-$1.5 & 2.5 \\

VNGP11 & 0.057 & 0.061 & 163.5 & 166.0 & 0.0 & 1.5 \\

VNGP12 & 0.054 & 0.061 & 206.0 & 208.0 & $-$1.5 & 0.5 \\

VNGP13 & 0.071 & 0.075 & 176.0 & 179.0 & $-$1.0 & 0.5 \\

VNGP14 & 0.079 & 0.094 & 186.5 & 204.2 & $-$7.5 & 0.5 \\

VNGP15 & 0.092 & 0.097 & 150.8 & 160.2 & $-$7.5 & $-$6.0 \\

VNGP16 & 0.093 & 0.100 & 147.0 & 149.5 & $-$6 & 1.5 \\

VNGP17 & 0.094 & 0.101 & 204.0 & 211.5 & $-$3.5 & 2.5 \\

VNGP18 & 0.099 & 0.115 & 147.5 & 153.6 & $-$4.0 & 1.0 \\

VNGP19 & 0.099 & 0.120 & 188.8 & 197.0 & $-$3.0 & 2.5 \\

VNGP20 & 0.112 & 0.120 & 176.6 & 178.0 & 0.5 & 2.5 \\

VNGP21 & 0.110 & 0.115 & 211.6 & 217.0 & $-$2.5 & $-$0.5 \\

VNGP22 & 0.110 & 0.123 & 200.2 & 205.0 & $-$3.25 & 2.5 \\

VNGP23 & 0.123 & 0.131 & 169.0 & 174.0 & $-$4.0 & $-$1.25 \\

VNGP24 & 0.127 & 0.131 & 182.5 & 188.0 & 0.0 & 2.5 \\

VNGP25 & 0.127 & 0.131 & 197.5 & 202.0 & $-$1.25 & 0.5 \\

VNGP26 & 0.134 & 0.142 & 179.5 & 187.5 & $-$2.0 & 2.5 \\

VNGP27 & 0.134 & 0.138 & 204.0 & 207.0 & $-$1.5 & 0.0 \\

VNGP28 & 0.138 & 0.150 & 147.0 & 154.6 & $-$3.5 & 2.5 \\

VNGP29 & 0.142 & 0.150 & 162.0 & 166.0 & $-$4.0 & 1.5 \\

VNGP30 & 0.142 & 0.150 & 172.6 & 178.0 & $-$1.0 & 2.5 \\

\\ \hline
\label{tab:ngpvoid}
\end{tabular}
\end{center}.
\end{table*}

\end{document}